# Large System Analysis of Base Station Cooperation for Power Minimization

Luca Sanguinetti, *Senior Member, IEEE*, Romain Couillet, *Senior Member, IEEE*, and Merouane Debbah, *Fellow, IEEE*

*Abstract*—This work focuses on a large-scale multi-cell multi-user MIMO system in which $L$ base stations (BSs) of $N$ antennas each communicate with $K$ single-antenna user equipments. We consider the design of the linear precoder that minimizes the total power consumption while ensuring target user rates. Three configurations with different degrees of cooperation among BSs are considered: the coordinated beamforming scheme (only channel state information is shared among BSs), the coordinated multipoint MIMO processing technology or network MIMO (channel state and data cooperation), and a single cell beamforming scheme (only local channel state information is used for beamforming while channel state cooperation is needed for power allocation). The analysis is conducted assuming that $N$ and $K$ grow large with $K/N$ non trivial ratio $K/N$ and imperfect channel state information (modeled by the generic Gauss-Markov formulation form) is available at the BSs. Tools of random matrix theory are used to compute, in explicit form, deterministic approximations for: (*i*) the parameters of the optimal precoder; (*ii*) the powers needed to ensure target rates; and (*iii*) the total transmit power. These results are instrumental to get further insight into the structure of the optimal precoders and also to reduce the implementation complexity in large-scale networks. Numerical results are used to validate the asymptotic analysis in the finite system regime and to make comparisons among the different configurations.

## I. Introduction

The road forward for satisfying the increasing number of users and high rate expectations in 5G systems is very high spatial utilization. Among the different technologies in this context, massive MIMO is considered as one of the most promising [1]–[4]. Under the assumption of uncorrelated channels, if the number of base station (BS) antennas $N$ goes to infinity and the number of user equipment terminals (UE) $K$ is maintained fixed, the performance of massive MIMO systems becomes limited only by the so-called pilot contamination and simple matched filter and maximum ratio transmission with no cooperation among cells can entirely eliminate the uplink and downlink multicell interference. However, the maximum number of antennas at each BS is limited in practice. In such a case, interference-aware precoder designs with cooperation among cells should be applied for optimal handling of the remaining intercell interference.

Coordinated multi-cell resource allocation is generally formulated as an optimization problem in which a desired network utility is maximized subject to some quality-of-service requirements. In this work, we focus on the problem of designing the optimal linear precoder in multi-cell networks for minimizing the total transmit power while ensuring a set of target user rates [5], [6]. This problem has received great attention in the last years [7] and is gaining renewed interest nowadays due to the emerging research area of green multi cellular networks [8]. We specifically consider the downlink of a multi-cell multi-user MIMO system in which $L$ BSs equipped with $N$ antennas each communicate with $K$ single-antenna UEs. Within this setting, several configurations with different degrees of cooperation can be envisioned [9]. In this work, the following three are considered: (*i*) the coordinated beamforming (CoBF) scheme in [10] in which each BS sends data to its own users only but channel state information (CSI) is shared between the $L$ BSs so that each BS can exploit its excess number of spatial dimensions to mitigate the interference generated in other cells; this has the advantage of not having to distribute all users' data to all BSs; (*ii*) the fully cooperative scheme, widely known in the literature as coordinated multipoint MIMO (CoMP) or also network MIMO [9], in which the BSs share the CSI of the UEs as well as their data through backhaul links[1]; and (*iii*) a single cell beamforming (ScBF) scheme in which each BS computes the beamformer only on the basis of the CSI of its own UEs while the power allocation requires the BSs to share CSI so that the rate constraints can be jointly satisfied. In all the above scenarios, under the assumption of perfect CSI, the optimal linear precoder is known to be a function of some Lagrange multipliers, the computation of which can be performed using convex optimization tools or solving a fixed-point problem [7]. Although numerically feasible, both approaches do not provide any insight into the structure of the optimal precoder. Moreover, the computation must be performed for any new realization of propagation channels, which is too computationally cumbersome when the network size becomes large (as envisioned in 5G networks).

L. Sanguinetti is with the University of Pisa, Dipartimento di Ingegneria dell'Informazione, Italy (luca.sanguinetti@unipi.it) and also with the Large Systems and Networks Group (LANEAS), CentraleSupélec, Université Paris-Saclay, 3 rue Joliot-Curie, 91192 Gif-sur-Yvette, France. R. Couillet is with the Telecommunication Department of CentraleSupélec, France. (romain.couillet@centralesupelec.fr). M. Debbah is the Large Systems and Networks Group (LANEAS), CentraleSupélec, Université Paris-Saclay, 3 rue Joliot-Curie, 91192 Gif-sur-Yvette, France (m.debbah@centralesupelec.fr) and also with the Mathematical and Algorithmic Sciences Lab, Huawei Technologies Co. Ltd., France (merouane.debbah@huawei.com).

This research has been supported by the ERC Starting Grant 305123 MORE, by the HUAWEI project RMTin5G and also by NEWCOM# (Grant agreement no. 318306). L. Sanguinetti is also funded by the People Programme (Marie Curie Actions) FP7 PIEF-GA-2012-330731 Dense4Green.

A preliminary version of this paper was presented at the IEEE Global Communication Conference, San Diego, USA, Dec. 2015.

---

[1]Although CoMP might refer to a more general setting in the 3GPP standard, in this work the term is used to refer to network MIMO.

To overcome these issues, we follow the same approach as in other works for single- or multi-cell networks [11]–[16] and resort to the asymptotic regime where $N$ and $K$ grow large with a non trivial ratio $K/N$. The design and analysis of the considered networks is performed under the assumption of imperfect CSI (modeled by the generic Gauss-Markov formulation form) for the UEs. Unlike most previous works [13], [16], the asymptotically optimal values of the Lagrange multipliers are computed using recent results from random matrix theory [17], which provide us with a much simpler means to overcome the technical difficulties arising with the application of standard random matrix theory tools. These results are then exploited to compute explicit expressions for the asymptotic signal-to-interference-plus-noise ratios (SINRs), which are eventually used to obtain the asymptotic powers needed to ensure target rates as well as the asymptotic total transmit power. As shall be seen, all the aforementioned deterministic approximations are found to depend only on the long-term channel attenuations of the UEs, the relative strength of interference among BSs, the target rates and the quality of the channel estimates. As a notable outcome of this work, the above analysis provides a unified framework that can be used to compare the considered networks under different settings and to eventually get insights on how the different parameters affect their performance. Moreover, in the same spirit of [18], [19], the provided results can be used to derive optimal distributed algorithms that rely only on the exchange, among nearby BSs, of long-term fading components. Numerical results are used to show that the asymptotic analysis well approximate the network performance in the finite system regime.

The main literature related to this work is represented by [12]–[16], [20]. Specifically, a single-cell setting is considered in [12], [20] while a CoBF network is investigated in [13]. Unlike [13], we provide closed-form expressions for the Lagrange multipliers, which are instrumental to also compute closed-form expressions for SINRs and transmit powers. In [15], the authors focus on the sum rate of a CoMP under the assumption of regularized zero-forcing precoding. In [16], the authors provide an asymptotic analysis of all considered network configurations but for the simplest case in which only two cells are present and CSI is perfect. The analysis is also conducted under the restrictive assumption that the channels among all UEs within a given cell and the interfering BS can be modelled as $\sim \mathcal{CN}(0, \epsilon \mathbf{I}_N)$ with $\epsilon$ controlling the interference level between neighbor cells. Moreover, the same rate is imposed to all UEs. We importantly show in the present article that, within our framework, there is no substantial additional difficulty in treating the more general setting of interest here which, unlike [16], considers multiple cells, different rate constraints, imperfect CSI and models the interference between neighbor cells according to the large scale fading coefficients.

The remainder of this paper is organized as follows. Next section describes the signal model and revisits the optimal linear precoder for the different network configurations. Section III deals with the large system analysis. In Section IV, numerical results are used to validate the theoretical analysis and to make comparisons among the different beamforming schemes. Finally, the major conclusions and implications are drawn in Section V. All the technical proofs are presented in the Appendices.

*Notation*: The following notation is used throughout the paper. Scalars are denoted by lower case letters whereas boldface lower (upper) case letters are used for vectors (matrices). The superscripts $^T$ and $^H$ denote transpose and conjugate transpose. We denote by $\mathbf{I}_N$ the identity matrix of order $N$ and call $\mathbf{0}_N$ and $\mathbf{1}_N$ the $N-$dimensional all-zero and all-one vectors, respectively. A random vector $\mathbf{x} \sim \mathcal{CN}(\mathbf{m}, \mathbf{C})$ is complex Gaussian distributed with mean $\mathbf{m}$ and covariance matrix $\mathbf{C}$. The notation $\mathrm{tr}(\mathbf{A})$ stands for the trace of matrix $\mathbf{A}$ whereas $\mathbf{A} = \mathrm{diag}\{x_1, \ldots, x_N\}$ denotes a diagonal matrix of order $N$. We use $[\cdot]_{i,k}$ to denote the $(i,k)$th element of the enclosed matrix and $\otimes$ to indicate the Kronecker product. We denote $a_n \asymp b_n$ the equivalence relation $a_n - b_n \to 0$ as $n \to \infty$ for two infinite sequences $a_n$ and $b_n$.

## II. System Model

Consider the downlink of a multi-cell multi-user MIMO system composed of $L$ cells, the BS of each cell comprising $N$ antennas to communicate with $K$ single-antenna UEs. As mentioned previously, we consider three different configurations with different degrees of cooperation: $i$) the coordinated beamforming scheme in [10]; $ii$) the coordinated multipoint processing (or network MIMO) [9]; and $iii$) a single cell beamforming scheme. In all these schemes, we are interested in minimizing the total transmit power $P_T$ while satisfying rate constraints at the UEs. Under the assumption of perfect CSI at the BSs, this problem can be solved using different approaches based for example on second order conic programming and standard decomposition techniques [7]. Next, we consider a finite size system and review the optimal precoder structure for the aforementioned schemes. In doing so, we assume that the feasibility conditions are satisfied [5]–[7], [13].

### A. Coordinated Beamforming

In the CoBF setting, each UE is attached to a specific serving BS while receiving interfering data from other BSs. As such, we shall use a double index notation to refer to each UE e.g., "user $k$ in cell $j$". Under this convention, let us thus define $\mathbf{h}_{ljk} \in \mathbb{C}^N$ as the channel from BS $l$ to UE $k$ in cell $j$, given by $\mathbf{h}_{ljk} = \sqrt{d_{ljk}}\mathbf{w}_{ljk}$ where $\mathbf{w}_{ljk} \in \mathbb{C}^N$ is the small-scale fading channel assumed to be Gaussian with zero mean and unit covariance, i.e., $\sim \mathcal{CN}(\mathbf{0}_N, \mathbf{I}_N)$, and $d_{ljk}$ accounts for the corresponding large scale channel fading or path loss (from BS $l$ to UE $k$ in cell $j$). Denoting by $\mathbf{g}_{jk} \in \mathbb{C}^N$ the precoding vector intended to UE $k$ in cell $j$, its received signal can be written as

$$y_{jk} = \mathbf{h}_{jjk}^H \mathbf{g}_{jk} s_{jk} + \sum_{i=1, i \neq k}^{K} \mathbf{h}_{jji}^H \mathbf{g}_{ji} s_{ji}$$
$$+ \sum_{l=1, l \neq j}^{L} \sum_{i=1}^{K} \mathbf{h}_{ljk}^H \mathbf{g}_{li} s_{li} + n_{jk} \quad (1)$$

with $s_{li} \in \mathbb{C}$ being the signal intended to user $i$ in cell $l$, assumed independent across $(l, i)$ pairs, of zero mean and unit variance, and $n_{jk} \sim \mathcal{CN}(0, \sigma^2)$. Under the assumption of target UE rates $\{r_{jk}; \forall j, k\}$, the power minimization problem for CoBF can be formulated as:

$$\min_{\{\mathbf{g}_{jk}\}} \quad \sum_{j=1}^{L} \sum_{k=1}^{K} \mathbf{g}_{jk}^H \mathbf{g}_{jk} \tag{2}$$

$$\text{s.t.} \quad \frac{|\mathbf{h}_{jjk}^H \mathbf{g}_{jk}|^2}{\sum_{i=1, i \neq j}^{K} |\mathbf{h}_{jjk}^H \mathbf{g}_{ji}|^2 + \sum_{l=1, l \neq j}^{L} \sum_{i=1}^{K} |\mathbf{h}_{ljk}^H \mathbf{g}_{li}|^2 + \sigma^2} \geq \gamma_{jk} \ \forall j, k$$

where $\gamma_{jk} = 2^{r_{jk}} - 1$ denote the corresponding SINR constraints. Upon existence [7], the unique solution of (2) is $\mathbf{g}_{jk}^\star = \sqrt{\frac{p_{jk}^\star}{N}} \frac{\mathbf{v}_{jk}^\star}{||\mathbf{v}_{jk}^\star||}$ with $\mathbf{v}_{jk}^\star = \left( \sum_{l=1}^{L} \sum_{k=1}^{K} \frac{\lambda_{li}^\star}{N} \mathbf{h}_{jli} \mathbf{h}_{jli}^H + \mathbf{I}_N \right)^{-1} \mathbf{h}_{jjk} \ \forall j, k$ where $\{\lambda_{jk}^\star / N\}$ are the Lagrange multipliers associated to the SINR constraints and are obtained as the unique fixed point solution of the following set of equations [5]–[7]:

$$\lambda_{jk}^\star = \frac{(1 + 1/\gamma_{jk})^{-1}}{\mathbf{h}_{jjk}^H \left( \sum_{l=1}^{L} \sum_{k=1}^{K} \lambda_{li}^\star \mathbf{h}_{jli} \mathbf{h}_{jli}^H + N \mathbf{I} N \right)^{-1} \mathbf{h}_{jjk}} \quad \forall j, k. \tag{3}$$

The optimal power values $\{p_{jk}^\star\}$ are such that the SINR constraints in (2) are all satisfied with equality. This amounts to computing the unique solution of the following set of equations [7]:

$$\frac{1}{\gamma_{jk}} \frac{p_{jk}}{N} \frac{|\mathbf{h}_{jjk}^H \mathbf{v}_{jk}^\star|^2}{||\mathbf{v}_{jk}^\star||^2} - \sum_{i=1, i \neq k}^{K} \frac{p_{ji}}{N} \frac{\left| \mathbf{h}_{jjk}^H \mathbf{v}_{ji}^\star \right|^2}{\left\| \mathbf{v}_{ji}^\star \right\|^2}$$

$$- \sum_{l=1, l \neq j}^{L} \sum_{i=1}^{K} \frac{p_{li}}{N} \frac{\left| \mathbf{h}_{ljk}^H \mathbf{v}_{li}^\star \right|^2}{\left\| \mathbf{v}_{li}^\star \right\|^2} + \sigma^2 = 0 \quad \forall j, k. \tag{4}$$

Observe also that the Lagrange multipliers $\{\lambda_{jk}^\star / N\}$ and vectors $\{\mathbf{v}_{jk}^\star\}$ may be thought of as the solution to the following dual UL power minimization problem [16]:

$$\min_{\{\mathbf{v}_{jk}, \lambda_{jk}\}} \quad \sum_{j=1}^{L} \sum_{k=1}^{K} \frac{\lambda_{jk}}{N} \sigma^2 \tag{5}$$

$$\text{s.t.} \quad \frac{\lambda_{jk} |\mathbf{v}_{jk}^H \mathbf{h}_{jjk}|^2}{\mathbf{v}_{jk}^H \left[ \sum_{(l,i) \neq (j,k)} \lambda_{li} \mathbf{h}_{jli} \mathbf{h}_{jli}^H + N \mathbf{I}_N \right] \mathbf{v}_{jk}} \geq \gamma_{jk} \ \forall j, k$$

with $\{\mathbf{v}_{jk}\}$ being the receive beamforming vectors.

### B. Coordinated Multipoint Processing

In the CoMP (or network MIMO) setting, each UE is jointly served by all BSs. In other words, there exists no cell-user association and thus the UEs can be indexed as $k$ from 1 to $KL$. Let us then denote $\mathbf{h}_k = [\mathbf{h}_{1k}^T, \ldots, \mathbf{h}_{L,k}^T]^T$ with $\mathbf{h}_{jk} \in \mathbb{C}^N$ the channel from BS $j$ to user $k$ given by $\mathbf{h}_{jk} = \sqrt{d_{jk}} \mathbf{w}_{jk}$ where $\mathbf{w}_{jk} \in \mathbb{C}^{NL}$ is the small-scale fading and $d_{jk}$ accounts for the path loss (from BS $j$ to UE $k$). Denoting by $\mathbf{g}_k \in \mathbb{C}^{NL}$

the joint precoding vector for UE $k$, its received signal can be written as

$$y_k = \mathbf{h}_k^H \mathbf{g}_k s_k + \sum_{i=1, i \neq k}^{KL} \mathbf{h}_k^H \mathbf{g}_i s_i + n_k \tag{6}$$

with $s_i \in \mathbb{C}$ being the signal intended to user $i$, independent across $i$, of zero mean and unit variance, and $n_k \sim \mathcal{CN}(0, \sigma^2)$. In the above setting, the power minimization problem takes the form:

$$\min_{\{\mathbf{g}_k\}} \quad \sum_{k=1}^{KL} \mathbf{g}_k^H \mathbf{g}_k \tag{7}$$

$$\text{s.t.} \quad \frac{|\mathbf{h}_k^H \mathbf{g}_k|^2}{\sum_{i=1, i \neq k}^{KL} |\mathbf{h}_k^H \mathbf{g}_i|^2 + \sigma^2} \geq \gamma_k \quad \forall k$$

where $\gamma_k = 2^{r_k} - 1$ with $r_k$ the rate constraint of UE $k$. The solution to (7) is $\mathbf{g}_k^\star = \sqrt{\frac{p_k^\star}{NL}} \frac{\mathbf{v}_k^\star}{||\mathbf{v}_k^\star||}$ with $\mathbf{v}_k^\star = \left( \sum_{i=1}^{KL} \frac{\lambda_i^\star}{NL} \mathbf{h}_i \mathbf{h}_i^H + \mathbf{I}_{NL} \right)^{-1} \mathbf{h}_k$ where $\{\lambda_i^\star / (NL)\}$ are such that [5]–[7]:

$$\lambda_k^\star = \frac{(1 + 1/\gamma_k)^{-1}}{\mathbf{h}_k^H \left( \sum_{i=1}^{KL} \lambda_i^\star \mathbf{h}_i \mathbf{h}_i^H + NL \mathbf{I}_{NL} \right)^{-1} \mathbf{h}_k} \quad \forall k. \tag{8}$$

As before, the optimal $\{p_k^\star\}$ are computed such that the SINR constraints in (7) are satisfied with equality [7]. Then, we obtain

$$\frac{1}{\gamma_k} \frac{p_k}{NL} \frac{|\mathbf{h}_k^H \mathbf{v}_k^\star|^2}{\|\mathbf{v}_k^\star\|^2} - \sum_{i=1, i \neq k}^{KL} \frac{p_i}{NL} \frac{|\mathbf{h}_i^H \mathbf{v}_i^\star|^2}{\|\mathbf{v}_i^\star\|^2} + \sigma^2 = 0 \quad \forall k. \tag{9}$$

As for CoBF, $\{\lambda_k^\star / (NL)\}$ and $\{\mathbf{v}_k^\star\}$ can be obtained as the solution to the following dual UL problem [16]:

$$\min_{\{\mathbf{v}_k, \lambda_k\}} \quad \sum_{k=1}^{KL} \frac{\lambda_k}{NL} \sigma^2 \tag{10}$$

$$\text{s.t.} \quad \frac{\lambda_k |\mathbf{v}_k^H \mathbf{h}_k|^2}{\mathbf{v}_k^H \left( \sum_{i \neq k} \lambda_i \mathbf{h}_i \mathbf{h}_i^H + NL \mathbf{I}_{NL} \right) \mathbf{v}_k} \geq \gamma_k \quad \forall k.$$

### C. Single Cell Beamforming

Inspired by a single cell beamforming scheme (see for example [12]), we also consider the following power minimization problem:

$$\min_{\{\mathbf{g}_{jk}\}} \quad \sum_{k=1}^{K} \mathbf{g}_{jk}^H \mathbf{g}_{jk} \tag{11}$$

$$\text{s.t.} \quad \frac{|\mathbf{h}_{jjk}^H \mathbf{g}_{jk}|^2}{\sum_{i=1, i \neq j}^{K} |\mathbf{h}_{jjk}^H \mathbf{g}_{ji}|^2 + \sum_{l=1, l \neq j}^{L} \sum_{i=1}^{K} |\mathbf{h}_{ljk}^H \mathbf{g}_{li}|^2 + \sigma^2} \geq \gamma_{jk} \ \forall k.$$

Upon existence, the solution to (11) is given by $\mathbf{g}_{jk}^\star = \sqrt{\frac{p_{jk}^\star}{N}} \frac{\mathbf{v}_{jk}^\star}{||\mathbf{v}_{jk}^\star||}$ with $\mathbf{v}_{jk}^\star = \left( \sum_{i=1}^{K} \frac{\lambda_{ji}^\star}{N} \mathbf{h}_{jji} \mathbf{h}_{jji}^H + \mathbf{I}_N \right)^{-1} \mathbf{h}_{jjk}$ where the scalars $\{\lambda_{jk}^{\star}/N\}$ are such that [5]–[7]:

$$\lambda_{jk}^{\star} = \frac{(1+1/\gamma_{jk})^{-1}}{\mathbf{h}_{jjk}^{H}\left(\sum_{i=1}^{K}\lambda_{ji}^{\star}\mathbf{h}_{jji}\mathbf{h}_{jji}^{H} + N\mathbf{I}_{N}\right)^{-1}\mathbf{h}_{jjk}} \quad \forall j,k. \quad (12)$$

The powers $\{p_{jk}^{\star}\}$ are computed such that the SINR constraints in (4) are all satisfied [7]. As for CoBF, this requires CSI to be exchanged among BSs. This makes the considered ScBF scheme different from a "classical" single cell processing technique in which each BS does not exchange any information with the other BSs.[2] The reason why it is referred to as a single cell scheme is due to the fact that, unlike CoBF, the computation of the beamforming vectors $\mathbf{v}_{jk}^{\star} = \left(\sum_{i=1}^{K}\frac{\lambda_{ji}^{\star}}{N}\mathbf{h}_{jji}\mathbf{h}_{jji}^{H} + \mathbf{I}_{N}\right)^{-1}\mathbf{h}_{jjk}$ of cell $j$ does not require knowledge of the fast fading channels $\{\mathbf{h}_{jli}; \forall l \neq j, i\}$ from all UEs in the other cells to cell $j$. More details on this will be given in Section III-C.

Similar to CoBF and CoMP, $\{\lambda_{jk}^{\star}/N\}$ and $\{\mathbf{v}_{jk}^{\star}\}$ may be thought of as the solution to the following dual UL power minimization problem for cell $j$:

$$\min_{\{\mathbf{v}_{jk},\lambda_{jk}\}} \quad \sum_{k=1}^{K}\frac{\lambda_{jk}}{N}\sigma_{jk}^{2} \quad (13)$$

$$\text{s.t.} \quad \frac{\lambda_{jk}|\mathbf{v}_{jk}^{H}\mathbf{h}_{jjk}|^{2}}{\mathbf{v}_{jk}^{H}\left[\sum_{i=1,i\neq k}^{K}\lambda_{ji}\mathbf{h}_{jji}\mathbf{h}_{jji}^{H} + N\mathbf{I}_{N}\right]\mathbf{v}_{jk}} \geq \gamma_{jk} \,\forall k$$

with $\{\mathbf{v}_{jk}\}$ being the receive beamforming vectors and $\sigma_{jk}^{2} = \sum_{l=1,l\neq j}^{L}\sum_{i=1}^{K}\lambda_{li}|\mathbf{v}_{jk}^{H}\mathbf{h}_{jli}|^{2} + \sigma^{2}$.

## III. Large System Analysis

Let $\boldsymbol{\lambda}^{\star}$ and $\mathbf{p}^{\star}$ denote for the three settings above the vectors collecting the Lagrange multipliers and power values, respectively. As shown previously, the precoding vectors are parameterized by $\boldsymbol{\lambda}^{\star}$ and $\mathbf{p}^{\star}$, where $\boldsymbol{\lambda}^{\star}$ needs to be evaluated by solving a set of fixed-point equations. This is a computationally demanding task when $N$ and $K$ are large since the matrix inversion operation in (3), (8) or (12) must be computed several times, with complexity proportional to $N^{2}KL$ or $(NL)^{2}KL$. Besides, from a practical standpoint, a close inspection of the SINR constraints in (4) and (9) reveals that the evaluation of $\mathbf{p}^{\star}$ requires knowledge of all channels $\{\mathbf{h}_{ljk}\}$ and $\{\mathbf{h}_{i}\}$, thus implying some channel exchange procedure within the network at the rate of the fast fading channel evolution. This makes the implementation of all the above solutions a difficult task, especially when $N, K$ become large. Finally, computing $\boldsymbol{\lambda}^{\star}$ as the fixed point of (3), (8) or (12) does not provide any insight into the optimal structure of $\boldsymbol{\lambda}^{\star}$ and $\mathbf{p}^{\star}$. To overcome these issues, we exploit the statistical distribution for $\mathbf{h}_{ljk}$ and $\mathbf{h}_{i}$ and the large values of $N, K$ to compute deterministic approximations of $\boldsymbol{\lambda}^{\star}$ and $\mathbf{p}^{\star}$ [21]. For

[2]This case is considered for example in [16] in which the SINR constraint is achieved by all UEs through an iterative procedure based on the bisection method. However, such a procedure can only be applied when the same SINR constraint is imposed to all UEs, which is not the case of the considered network.

technical purposes, we assume the following grow rate of the system dimensions:

**Assumption 1.** *As* $N \rightarrow \infty$, $0 \leq \liminf_{N\rightarrow\infty} K/N \leq \limsup_{N\rightarrow\infty} K/N < \infty$.

A known problem with the asymptotic analysis is that the target rates are not guaranteed to be achieved when $N$ is finite and relatively small (e.g., [14]). This is because the approximation errors translate into fluctuations of the resulting SINR values. However, these errors vanish rapidly when $N$ takes large yet finite values as it is envisioned for massive MIMO systems [2].

We further assume that only imperfect CSI is available at the BSs. Since the optimal linear precoder for the aforementioned network configurations is not known when only imperfect CSI is available, we overcome this issue by simply replacing the true channels with their estimates (which is an accurate procedure for good CSI quality).

### A. Coordinated Beamforming

Let $\widehat{\mathbf{h}}_{ljk}$ be an estimate of $\mathbf{h}_{ljk}$ and assume, similar to [11] (among many others), that this can be modeled by the generic Gauss-Markov formulation:

$$\widehat{\mathbf{h}}_{ljk} = \sqrt{d_{ljk}}\left(\sqrt{1-\tau_{ljk}^{2}}\mathbf{w}_{ljk} + \tau_{ljk}\mathbf{q}_{ljk}\right) = \sqrt{d_{ljk}}\mathbf{z}_{ljk} \quad (14)$$

where $\mathbf{q}_{ljk} \sim \mathcal{CN}(0, \mathbf{I}_{N})$ accounts for the channel estimation errors independent of $\mathbf{w}_{ljk}$ and $\mathbf{z}_{ljk} = \sqrt{1-\tau_{ljk}^{2}}\mathbf{w}_{ljk} + \tau_{ljk}\mathbf{q}_{ljk}$. The parameter $\tau_{ljk} \in [0,1]$ reflects the accuracy or quality of the channel estimate, i.e., $\tau_{ljk} = 0$ for perfect CSI and $\tau_{ljk} = 1$ for a channel estimate completely independent of the genuine channel. Replacing $\{\mathbf{h}_{jli}\}$ by $\{\widehat{\mathbf{h}}_{jli}\}$ into (3) yields

$$\lambda_{jk}^{\star} = \frac{(1+1/\gamma_{jk})^{-1}}{\widehat{\mathbf{h}}_{jjk}^{H}\left(\sum_{l=1}^{L}\sum_{i=1}^{K}\lambda_{li}^{\star}\widehat{\mathbf{h}}_{jli}\widehat{\mathbf{h}}_{jli}^{H} + N\mathbf{I}_{N}\right)^{-1}\widehat{\mathbf{h}}_{jjk}} \quad \forall j,k. \quad (15)$$

As mentioned earlier, the implicit formulation for $\lambda_{jk}^{\star}$ prevents any insightful analysis of the system performance. By a large dimensional analysis, exploiting recent tools from random matrix theory (see notably [17]), we shall subsequently show that $\lambda_{jk}^{\star}$ gets asymptotically close to an explicit deterministic quantity as $N$ and $K$ grow large as for Assumption 1, and that this quantity provides clear insight on the behavior of the precoder and the system as a whole.

For technical reasons, the following reasonable assumption is imposed on the system settings.

**Assumption 2.** *The* $\{d_{ijk}\}$ *and* $\{\gamma_{jk}\}$ *satisfy* $\limsup\max\{d_{ijk}\} < \infty$ *and* $\limsup\max\{\gamma_{jk}\} < \infty$.

A main technical result then lies in the following theorem:

**Theorem 1.** *Let Assumptions 1 and 2 hold. Then,* $\max_{jk}|\lambda_{jk}^{\star} - \overline{\lambda}_{jk}^{(CoBF)}| \rightarrow 0$ *almost surely with*

$$\overline{\lambda}_{jk}^{(CoBF)} = \frac{1}{\eta_{j}}\frac{\gamma_{jk}}{d_{jjk}} \quad (16)$$

where $\{\eta_j\}$ is the unique positive solution to the following set of equations

$$\eta_j = \left( \frac{1}{N} \sum_{l=1}^{L} \sum_{i=1}^{K} \frac{\gamma_{li} \frac{d_{jli}}{d_{lli}} \frac{1}{\eta_l}}{1 + \gamma_{li} \frac{d_{jli}}{d_{lli}} \frac{\eta_j}{\eta_l}} + 1 \right)^{-1} \quad \forall j \qquad (17)$$

or, equivalently,

$$\eta_j = 1 - \frac{1}{N} \sum_{l=1}^{L} \sum_{i=1}^{K} \frac{\gamma_{li} \frac{d_{jli}}{d_{lli}} \frac{\eta_j}{\eta_l}}{1 + \gamma_{li} \frac{d_{jli}}{d_{lli}} \frac{\eta_j}{\eta_l}} \quad \forall j. \qquad (18)$$

*Proof:* The main difficulty lies in the implicit definition of the $\lambda_{jk}^\star$'s in (15). A first step consists in heuristically discarding the implicit structure to retrieve the expression for $\overline{\lambda}_{jk}^{(CoBF)}$ in explicit form. To proceed with an accurate proof, in Appendix A we follow similar steps as in [17] (in a completely different context though), by controlling the ratio $\overline{\lambda}_{jk}^{(CoBF)} / \lambda_{jk}^\star$. ∎

Some important insights can be readily extracted from Theorem 1. To begin with, observe that the computation of $\{\overline{\lambda}_{jk}^{(CoBF)}\}$ in (16) for cell $j$ requires only the knowledge of the SINR constraints $\{\gamma_{li}; \forall l, i\}$ and the average channel attenuations $\{d_{jli}; \forall l, i\}$ from all UEs to BS $j$. Differently from the fast fading channels $\{\mathbf{h}_{jli}\}$ required in (15), the latter can be accurately estimated as they change slowly with time (relative to the small-scale fading). Also, the Lagrange multiplier $\lambda_{jk}$ is known to act as a user priority parameter that implicitly determines how much interference a specific UE $k$ in cell $j$ may induce to the other UEs [7]. From (16), it turns out that $\overline{\lambda}_{jk}^{(CoBF)}$ is proportional to $\gamma_{jk}$ and inversely proportional to $d_{jjk}$ such that higher priority is given to UEs that require high performance or have weak propagation conditions. Moreover, observe that (18) can be equivalently rewritten as:

$$\eta_j = \varsigma_j - \frac{1}{N} \sum_{l=1, l \neq j}^{L} \sum_{i=1}^{K} \frac{\gamma_{li} \frac{d_{jli}}{d_{lli}} \frac{\eta_j}{\eta_l}}{1 + \gamma_{li} \frac{d_{jli}}{d_{lli}} \frac{\eta_j}{\eta_l}} \quad \forall j \qquad (19)$$

where $\varsigma_j$ is defined as

$$\varsigma_j \triangleq 1 - \frac{1}{N} \sum_{k=1}^{K} \frac{\gamma_{jk}}{1 + \gamma_{jk}}. \qquad (20)$$

From the above equation, it follows that $\eta_j$ in cell $j$ is such that $\eta_j \leq \varsigma_j$ and also it depends not only on its own cell requirements $\{\gamma_{jk}\}$ through $\varsigma_j$ but also on all the other cells through the ratios $\{\gamma_{li} \frac{d_{jli}}{d_{lli}} \frac{\eta_j}{\eta_l}\}$ and might be thought of as a cell priority parameter: higher priority is given to cell $j$ if $\eta_j$ is small. In particular, $d_{jli}/d_{lli}$ describes the relative strength of the interference received at UE $i$ in cell $l$ from BS $j$; for a given set of $\{\eta_{li}\}$ it is almost one for cell edge UEs of neighboring cells, while it is almost zero when cell $l$ is very distant from BS $j$. In other words, higher priority is given to those cells that create high interference, as it should. To get further insights on this, please refer to the simple two-cell two-user network case study considered in Section III-E.

If a completely symmetric scenario is considered, i.e., $d_{jli}/d_{lli} = d_{lji}/d_{jji}$ and $\gamma_{li} = \gamma_i \ \forall j, l$, the coefficients $\{\eta_j\}$ are all equal to a given $\eta$, which can be computed in explicit form as stated in the following corollary (similar results were obtained in [16] for a two-cell only network):

**Corollary 1.** *If a completely symmetric scenario is considered, then $\forall j$ $\eta_j = \eta$ with*

$$\eta = 1 - \frac{1}{N} \sum_{l=1}^{L} \sum_{i=1}^{K} \frac{\gamma_i \frac{d_{jli}}{d_{lli}}}{1 + \gamma_i \frac{d_{jli}}{d_{lli}}}. \qquad (21)$$

*Proof:* The proof follows directly from Theorem 1 using $d_{jli} = d_{lji}$ and $\gamma_{li} = \gamma_i \ \forall j, l$. ∎

An explicit form for $\{\eta_j\}$ can also be obtained in the high SINR regime as stated below:

**Corollary 2.** *If $\forall l, i \ \gamma_{li}$ grows large, then $\forall j$ $\lim_{\forall l, i \gamma_{li} \to \infty} \eta_j = \eta$ with $\eta = 1 - KL/N$ provided that $1 - KL/N > 0$.*

*Proof:* The proof follows directly from (17) or (18) of Theorem 1 observing that for given sets of $\{d_{jli}\}$ and $\{d_{lli}\}$ the ratio $\frac{\gamma_{li} \frac{d_{jli}}{d_{lli}} \frac{\eta_j}{\eta_l}}{1 + \gamma_{li} \frac{d_{jli}}{d_{lli}} \frac{\eta_j}{\eta_l}} \to 1$ as $\forall l, i \ \gamma_{li} \to \infty$. ∎

We now proceed to computing the asymptotic powers $\{\overline{p}_{jk}^{(CoBF)}\}$ satisfying the SINR constraints in the large-$(N, K)$ regime when imperfect CSI is available. To this end, we first compute the asymptotic values of the SINRs under the assumption that the transmit powers $\{p_{jk}\}$ are held fixed at a constant value and that $\mathbf{g}_{jk}$ is replaced with $\mathbf{g}_{jk} = \sqrt{\frac{p_{jk}}{N}} \frac{\hat{\mathbf{v}}_{jk}^\star}{\|\hat{\mathbf{v}}_{jk}^\star\|}$ such that

$$\text{SINR}_{jk} = \frac{\frac{p_{jk}}{N} \frac{|\mathbf{h}_{jjk}^H \hat{\mathbf{v}}_{jk}^\star|^2}{\|\hat{\mathbf{v}}_{jk}^\star\|^2}}{\sum_{i=1, i \neq k}^{K} \frac{p_{ji}}{N} \frac{|\mathbf{h}_{jjk}^H \hat{\mathbf{v}}_{ji}^\star|^2}{\|\hat{\mathbf{v}}_{ji}^\star\|^2} + \sum_{l=1, l \neq j}^{L} \sum_{i=1}^{K} \frac{p_{li}}{N} \frac{|\mathbf{h}_{ljk}^H \hat{\mathbf{v}}_{li}^\star|^2}{\|\hat{\mathbf{v}}_{li}^\star\|^2} + \sigma^2} \qquad (22)$$

where $\hat{\mathbf{v}}_{jk}^\star = \left( \sum_{l=1}^{L} \sum_{i=1}^{K} \frac{\lambda_{li}^\star}{N} \hat{\mathbf{h}}_{jli} \hat{\mathbf{h}}_{jli}^H + \mathbf{I}_N \right)^{-1} \hat{\mathbf{h}}_{jjk}$. We then have the following result:

**Lemma 1.** *Under Assumptions 1 and 2, $\max_{jk} |\text{SINR}_{jk} - \overline{\text{SINR}}_{jk}^{(CoBF)}| \to 0$ almost surely with*

$$\overline{\text{SINR}}_{jk}^{(CoBF)} = \frac{p_{jk} d_{jjk} (1 - \tau_{jjk}^2) \frac{1 - \frac{1}{N} \sum_{l=1}^{L} \sum_{i=1}^{K} \frac{\left( \gamma_{li} \frac{d_{jli}}{d_{lli}} \frac{\eta_j}{\eta_l} \right)^2}{\left( 1 + \gamma_{li} \frac{d_{jli}}{d_{lli}} \frac{\eta_j}{\eta_l} \right)^2}}{\overline{I}_{jk}^{(CoBF)} + \sigma^2}} \qquad (23)$$

where $\overline{I}_{jk}^{(CoBF)} \triangleq \sum_{l=1}^{L} \beta_{ljk} \left( \frac{1}{N} \sum_{i=1}^{K} p_{li} \right)$ with

$$\beta_{ljk} \triangleq d_{ljk} \frac{1 - \tau_{ljk}^2 \left[ 1 - \left( 1 + \gamma_{jk} \frac{d_{ljk}}{d_{jjk}} \frac{\eta_j}{\eta_l} \right)^2 \right]}{\left( 1 + \gamma_{jk} \frac{d_{ljk}}{d_{jjk}} \frac{\eta_j}{\eta_l} \right)^2}. \qquad (24)$$

*Proof:* Substituting $\overline{\lambda}_{jk}^{(CoBF)}$ for $\lambda_{jk}^\star$, the result follows as shown in Appendix B. ∎

For notational convenience, let us now denote by $\mathbf{b} = [b_1, \ldots, b_L]^T$ the vector with entries

$$b_j \triangleq \frac{1}{N} \sum_{i=1}^{K} \frac{\gamma_{ji}}{d_{jji}(1 - \tau_{jji}^2)}. \qquad (25)$$

The main result of this section unfolds from the previous lemma and provides the transmit power dedicated to each user and the minimal total transmit power to meet the SINR constraints.

**Theorem 2.** *Let $\boldsymbol{\Gamma} \in \mathbb{C}^{L \times L}$ be diagonal with entries*

$$[\boldsymbol{\Gamma}]_{j,j} \triangleq 1 - \frac{1}{N} \sum_{l=1}^{L} \sum_{i=1}^{K} \frac{\left(\gamma_{li} \frac{d_{jli}}{d_{lli}} \frac{\eta_j}{\eta_l}\right)^2}{\left(1 + \gamma_{li} \frac{d_{jli}}{d_{lli}} \frac{\eta_j}{\eta_l}\right)^2} \quad (26)$$

*and $\mathbf{F} \in \mathbb{C}^{L \times L}$ such that*

$$[\mathbf{F}]_{j,l} \triangleq \frac{1}{N} \sum_{k=1}^{K} \frac{\gamma_{jk} \beta_{ljk}}{d_{jjk}(1 - \tau_{jjk}^2)}. \quad (27)$$

*where $\beta_{ljk}$ is defined in (24). If and only if $\limsup_K \|\boldsymbol{\Gamma}^{-1}\mathbf{F}\| < 1$, then under Assumptions 1 and 2, the powers $\{p_{jk}\}$ required to meet the SINR constraints in the asymptotic regime are obtained as:*

$$\overline{p}_{jk}^{(CoBF)} = \frac{\gamma_{jk}}{d_{jjk}\left(1 - \tau_{jjk}^2\right)} \frac{\sum_{l=1}^{L} \beta_{ljk} \overline{P}_l^{(CoBF)} + \sigma^2}{1 - \frac{1}{N}\sum_{l=1}^{L}\sum_{i=1}^{K} \frac{\left(\gamma_{li}\frac{d_{jli}}{d_{lli}}\frac{\eta_j}{\eta_l}\right)^2}{\left(1+\gamma_{li}\frac{d_{jli}}{d_{lli}}\frac{\eta_j}{\eta_l}\right)^2}} \quad (28)$$

*where $\overline{\mathbf{P}}^{(CoBF)} = [\overline{P}_1^{(CoBF)}, \ldots, \overline{P}_L^{(CoBF)}]^T = \sigma^2(\boldsymbol{\Gamma}-\mathbf{F})^{-1}\mathbf{b}$ collects the total transmit power of each BS with $\mathbf{b}$ defined as in (25). Moreover, the asymptotic total transmit power is given by*

$$\overline{P}_T^{(CoBF)} = \mathbf{1}^T \overline{\mathbf{P}}^{(CoBF)} = \sigma^2 \mathbf{1}^T (\boldsymbol{\Gamma} - \mathbf{F})^{-1} \mathbf{b}. \quad (29)$$

*Proof:* The proof is given in Appendix C and basically proceeds as follows. The transmit powers are set to ensure that the SINR constraints are reached exactly in the asymptotic regime, that is such that $\overline{\text{SINR}}_{jk}^{(CoBF)} = \gamma_{jk}$ (with $\overline{\text{SINR}}_{jk}^{(CoBF)}$ defined in (23)). It then suffices to solve the implicit equation $\overline{\text{SINR}}_{jk}^{(CoBF)} = \gamma_{jk}$ in the unknowns $\{p_{jk}^{(CoBF)}\}$. This equation turns out to unwrap as an explicit equation for the $\{\overline{p}_{jk}^{(CoBF)}\}$, which are then readily obtained as in the theorem statement. The asymptotic approximation of the total transmit power easily follows taking into account that the transmit power of BS $j$ is given by $\frac{1}{N}\sum_{k=1}^{K} p_{jk}^{(CoBF)} = P_j^{(CoBF)}$. ∎

A close inspection of Theorem 2 reveals that the computation of $\{\overline{p}_{jk}^{(CoBF)}\}$ only requires knowledge of the system parameters $\{\gamma_{li}\}$ and $\{\tau_{li}\}$ and the large scale channel fading components $\{d_{jlk}\}$. The latter change slowly in time compared to small-scale fading components and can be accurately estimated and possibly exchanged among coupled BSs with a reasonable effort. This is in sharp contrast to the finite system regime wherein the evaluation of $\{p_{jk}^\star\}$ through (4) requires some channel exchange procedure within the network at the rate of the fast fading channel evolution.

From [22, Corollary 3], we have that $\limsup_K \|\boldsymbol{\Gamma}^{-1}\mathbf{F}\| < 1$ if none of the column sums of $\boldsymbol{\Gamma}^{-1}\mathbf{F}$ exceed unity and at least one is less than unity. This amounts to checking that $\sum_{l=1}^{L} \left[\boldsymbol{\Gamma}^{-1}\mathbf{F}\right]_{j,l} \leq 1 \ \forall l$ or, equivalently, (since $\boldsymbol{\Gamma}$ is diagonal)

$\sum_{l=1}^{L} [\mathbf{F}]_{j,l} \leq [\boldsymbol{\Gamma}]_{j,j} \ \forall j$ with the strict inequality holding for at least one $j$. Therefore, it follows that if $\forall j \in \{1, \ldots, L\}$

$$\frac{1}{N}\sum_{l=1}^{L}\sum_{k=1}^{K} \frac{\gamma_{jk}\beta_{ljk}}{d_{jjk}(1-\tau_{jjk}^2)} \leq$$
$$1 - \frac{1}{N}\sum_{l=1}^{L}\sum_{i=1}^{K} \frac{\left(\gamma_{li}\frac{d_{jli}}{d_{lli}}\frac{\eta_j}{\eta_l}\right)^2}{\left(1+\gamma_{li}\frac{d_{jli}}{d_{lli}}\frac{\eta_j}{\eta_l}\right)^2} \quad (30)$$

with the strict inequality holding for at least one $j$, then $\limsup_K \|\boldsymbol{\Gamma}^{-1}\mathbf{F}\| < 1$. Assume now that the quality of the channel estimates is the same for all UEs, i.e., $\tau_{ljk} = \tau_j \ \forall l, k$, then from the above discussion and using (24) (after simple calculus) the maximum level of imperfect CSI in cell $j$ is found to be:

**Lemma 2.** *Assume $\tau_{ljk} = \tau_j \ \forall l, k$. Then, for any given set of $\{\gamma_{jk}\}, \{d_{jjk}\}$ and $\{d_{ljk}\}$ a feasible asymptotic power allocation exists in the CoBF case if $\tau_j < \tau_{j,\max}^{(CoBF)}$ with*

$$\tau_{j,\max}^{(CoBF)} = \left(1 + \frac{\frac{1}{N}\sum_{k=1}^{K}\frac{\gamma_{jk}}{d_{jjk}}\sum_{l=1}^{L}d_{jlk}}{1 - \frac{1}{N}\sum_{l=1}^{L}\sum_{i=1}^{K}\frac{\left(\gamma_{li}\frac{d_{jli}}{d_{lli}}\frac{\eta_j}{\eta_l}\right)^2 + \gamma_{li}\frac{d_{jli}}{d_{lli}}}{\left(1+\gamma_{li}\frac{d_{jli}}{d_{lli}}\frac{\eta_j}{\eta_l}\right)^2}}\right)^{-1/2}. \quad (31)$$

To confirm the asymptotic optimality of the power allocation provided by Theorem 2, we now provide the following corollary that proves that the uplink-downlink duality holds true also in the asymptotic regime when perfect knowledge of the channel is available:

**Corollary 3.** *Under Assumptions 1, 2 and 4, if perfect knowledge of the channel is available, i.e., $\tau_{ljk} = 0 \ \forall l, j, k$, then the duality gap between (2) and (5) is zero.*

*Proof:* The proof is sketched in Appendix D. ∎

### B. Coordinated Multipoint Processing

With a slight abuse of notation, let $\widehat{\mathbf{h}}_{jk} \in \mathbb{C}^{NL}$ be the estimate of the channel from BS $j$ to user $k$ given by $\widehat{\mathbf{h}}_{jk} = \sqrt{d_{jk}}\left(\sqrt{1-\tau_k^2}\mathbf{w}_{jk} + \tau_k\mathbf{q}_{jk}\right) = \sqrt{d_{jk}}\mathbf{z}_{jk}$ where $\mathbf{q}_{jk} \sim \mathcal{CN}(\mathbf{0}, \mathbf{I}_N)$ and $\mathbf{z}_{jk} = \sqrt{1-\tau_k^2}\mathbf{w}_{jk} + \tau_k\mathbf{q}_{jk}$. Then, we may write $\widehat{\mathbf{h}}_k = [\widehat{\mathbf{h}}_{1k}^T, \ldots, \widehat{\mathbf{h}}_{L,k}^T]^T$ as

$$\widehat{\mathbf{h}}_k = \boldsymbol{\Theta}_k^{1/2}\left(\sqrt{1-\tau_k^2}\mathbf{w}_k + \tau_k\mathbf{q}_k\right) = \boldsymbol{\Theta}_k^{1/2}\mathbf{z}_k \quad (32)$$

with $\boldsymbol{\Theta}_k \in \mathbb{C}^{NL \times NL}$ being defined as $\boldsymbol{\Theta}_k = \text{diag}\{d_{1k}, \ldots, d_{Lk}\} \otimes \mathbf{I}_N$ and $\mathbf{q}_k = [\mathbf{q}_{1k}^T, \ldots, \mathbf{q}_{Lk}^T]^T$ and $\mathbf{w}_k = [\mathbf{w}_{1k}^T, \ldots, \mathbf{w}_{Lk}^T]^T$. Replacing $\mathbf{h}_k$ with $\widehat{\mathbf{h}}_k$ in (32) yields

$$\lambda_k^\star = \frac{(1 + 1/\gamma_k)^{-1}}{\widehat{\mathbf{h}}_k^H \left(\sum_{i}^{KL} \lambda_i^\star \widehat{\mathbf{h}}_i \widehat{\mathbf{h}}_i^H + NL\mathbf{I}_{NL}\right)^{-1}\widehat{\mathbf{h}}_k} \quad (33)$$

$\forall k \in \{1, 2, \ldots, KL\}$. Similar to the previous network configuration, we shall require here the following technical setting.

**Assumption 3.** *The $\{d_{jk}\}$ and $\{\gamma_k\}$ satisfy $\limsup \max_{j,k} d_{jk} < \infty$ and $\limsup \max_k \gamma_k < \infty$.*

Our first result in this setting is as follows:

**Theorem 3.** *Under Assumptions 1 and 3, $\max_k |\lambda_k^\star - \overline{\lambda}_k^{(CoMP)}| \to 0$ almost surely with*

$$\overline{\lambda}_k^{(CoMP)} = \frac{\gamma_k}{\frac{1}{L}\sum\limits_{l=1}^{L} d_{lk}\mu_l} \tag{34}$$

*where $\{\mu_l\}$ is the unique positive solution to the following set of equations:*

$$\mu_l = \left(\frac{1}{NL}\sum_{i=1}^{KL} \frac{d_{li}}{\frac{1}{L}\sum\limits_{j=1}^{L} d_{ji}\mu_j} \frac{\gamma_i}{1+\gamma_i} + 1\right)^{-1} \quad \forall l. \tag{35}$$

*Proof:* The proof is given in Appendix E and proceeds as that of Theorem 1. ∎

Unlike (16), in the CoMP configuration the Lagrange multiplier of UE $k$ is found to be inversely proportional to a weighted priority parameter given by

$$\epsilon_k = \frac{1}{L}\sum_{l=1}^{L} d_{lk}\mu_l \tag{36}$$

which basically takes into account the effort of each cell for jointly serving user $k$ – see Section III-E for an illustrative example. Replacing $\mathbf{g}_k$ with $\widehat{\mathbf{g}}_k = \sqrt{\frac{p_k}{NL}} \frac{\widehat{\mathbf{v}}_k^\star}{||\widehat{\mathbf{v}}_k^\star||}$, the SINR of user $k$ takes the form

$$\text{SINR}_k = \frac{\frac{p_k}{NL} \frac{|\mathbf{h}_k^H \widehat{\mathbf{v}}_k^\star|^2}{||\widehat{\mathbf{v}}_k^\star||^2}}{\sum\limits_{i=1, i\neq k}^{KL} \frac{p_i}{NL} \frac{|\mathbf{h}_k^H \widehat{\mathbf{v}}_i^\star|^2}{||\widehat{\mathbf{v}}_i^\star||^2} + \sigma^2} \tag{37}$$

where $\widehat{\mathbf{v}}_k^\star = \left(\sum_{i=1}^{KL} \frac{\lambda_i^\star}{NL}\widehat{\mathbf{h}}_i\widehat{\mathbf{h}}_i^H + \mathbf{I}_{NL}\right)^{-1}\widehat{\mathbf{h}}_k$. To proceed further, we call $\epsilon_k' = [\epsilon_{1k}', \ldots, \epsilon_{KLk}']^T$ the vector obtained as $\epsilon_k' = (\mathbf{I}_{KL} - \mathbf{J})^{-1}\mathbf{c}_k'$ where $\mathbf{c}_k' \in \mathbb{C}^{KL}$ is such that

$$[\mathbf{c}_k']_i = \frac{\gamma_k^2}{\left(\frac{1}{L}\sum\limits_{l=1}^{L} d_{lk}\mu_l\right)^2} \left(\frac{1}{L}\sum_{l=1}^{L} d_{li}d_{lk}\mu_l^2\right) \tag{38}$$

and $\mathbf{J} \in \mathbb{C}^{KL \times KL}$ has entries given by $[\mathbf{J}]_{i,k} = \frac{[\mathbf{c}_k']_i}{NL(1+\gamma_k)^2}$. Our main results are then as follows:

**Lemma 3.** *Under Assumptions 1 and 3, $\max_k |\text{SINR}_k - \overline{\text{SINR}}_k^{(CoMP)}| \to 0$ almost surely with*

$$\overline{\text{SINR}}_k^{(CoMP)} = p_k \frac{\epsilon_k^2}{\epsilon_k'} \frac{1-\tau_k^2}{\overline{T}_k^{(CoMP)} + \sigma^2} \tag{39}$$

*where $\overline{T}_k^{(CoMP)} \triangleq \frac{1-\tau_k^2\left[1-(1+\gamma_k)^2\right]}{(1+\gamma_k)^2}\left(\frac{1}{NL}\sum_{i=1}^{KL} p_i \frac{\epsilon_{ik}'}{\epsilon_i'}\right)$ and $\epsilon' = [\epsilon_1', \ldots, \epsilon_{KL}']^T = (\mathbf{I}_{KL} - \mathbf{J})^{-1}\mathbf{c}$ where $\mathbf{c} \in \mathbb{C}^{KL}$ has elements $[\mathbf{c}]_i = \frac{1}{L}\sum_{l=1}^{L} d_{lk}\mu_l^2$.*

*Proof:* See Appendix F. ∎

**Theorem 4.** *Let $\mathbf{Z} \in \mathbb{C}^{KL \times KL}$ be such that*

$$[\mathbf{Z}]_{k,i} \triangleq \frac{1}{NL} \frac{\gamma_i}{1-\tau_i^2} \frac{\epsilon_{ik}'}{\epsilon_i^2} \frac{1-\tau_i^2\left[1-(1+\gamma_i)^2\right]}{(1+\gamma_i)^2}. \tag{40}$$

*If and only if $\limsup_K ||\mathbf{Z}|| < 1$, then under Assumptions 1 and 3, the powers $\{p_k\}$ required to meet the SINR constraints in the asymptotic regime are obtained as:*

$$\overline{p}_k^{(CoMP)} = \frac{\gamma_k}{1-\tau_k^2} \frac{\epsilon_k'}{\epsilon_k^2}\left(\overline{\Omega}_k \frac{1-\tau_k^2\left[1-(1+\gamma_k)^2\right]}{(1+\gamma_k)^2} + \sigma^2\right) \tag{41}$$

*where $\overline{\Omega} = [\overline{\Omega}_1, \overline{\Omega}_2, \ldots, \overline{\Omega}_{KL}]^T$ is computed as $\overline{\Omega} \triangleq \sigma^2(\mathbf{I}_{KL} - \mathbf{Z})^{-1}\mathbf{z}$ and $\mathbf{z} \in \mathbb{C}^{KL}$ with $z_k \triangleq \frac{1}{NL}\sum_{i=1}^{KL} \frac{\gamma_i}{1-\tau_i^2} \frac{\epsilon_{ik}'}{\epsilon_i^2}$. Moreover, the asymptotic total transmit power is*

$$\overline{P}_T^{(CoMP)} = \frac{1}{NL}\sum_{i=1}^{KL} \overline{p}_i^{(CoMP)}. \tag{42}$$

*Proof:* The proof follows from Lemma 3 using the same arguments for proving Theorem 2. ∎

As done before for both CoBF, we observe that if

$$\frac{1}{NL}\sum_{k=1}^{KL} \frac{\gamma_i}{1-\tau_i^2} \frac{\epsilon_i'}{\epsilon_i^2} \frac{1-\tau_i^2\left[1-(1+\gamma_i)^2\right]}{(1+\gamma_i)^2} \leq 1 \tag{43}$$

and less than unity for at least one $k$, then $\limsup_K ||\mathbf{Z}|| < 1$. Therefore, we have that:

**Lemma 4.** *For any given set of $\{\gamma_i\}$ and $\{d_{li}\}$, a feasible power allocation exists for CoMP if $\tau_i < \tau_{i,\max}^{(CoMP)}$ with*

$$\tau_{i,\max}^{(CoMP)} = \left(1 + \frac{\gamma_i}{\epsilon_i^2(1+\gamma_i)^2} \frac{\frac{1}{NL}\sum\limits_{k=1}^{KL} \epsilon_{ik}'}{1 - \frac{1}{NL}\frac{\gamma_i}{\epsilon_i^2}\sum\limits_{k=1}^{KL} \epsilon_{ik}'}\right)^{-1/2}. \tag{44}$$

**Corollary 4.** *Under Assumptions 1 and 3, if perfect knowledge of the channel is available, i.e., $\tau_k = 0$ $\forall k$, then the duality gap between (7) and (10) is zero.*

*Proof:* Despite being much more involved, the proof basically unfolds from the same arguments used for proving Corollary 2. ∎

### C. Single Cell Beamforming

Replacing $\mathbf{h}_{ljk}$ with $\widehat{\mathbf{h}}_{ljk}$ into (12) yields:

$$\lambda_{jk}^\star = \frac{(1+1/\gamma_{jk})^{-1}}{\widehat{\mathbf{h}}_{jjk}^H\left(\sum\limits_{i=1}^{K}\lambda_{ji}^\star\widehat{\mathbf{h}}_{jji}\widehat{\mathbf{h}}_{jji}^H + N\mathbf{I}_N\right)^{-1}\widehat{\mathbf{h}}_{jjk}} \quad \forall j, k. \tag{45}$$

**Assumption 4.** *The $\{\gamma_{jk}\}$ satisfy $\limsup_N \frac{1}{N}\sum_{i=1}^{K} \frac{\gamma_{ji}}{1+\gamma_{ji}} < 1$.*

When the Lagrange multipliers are computed as above, we have that:

**Theorem 5.** *Let Assumptions 1, 2 and 4 hold. Then, $\max_{jk} |\lambda_{jk}^\star - \overline{\lambda}_{jk}^{(ScBF)}| \to 0$ almost surely with*

$$\overline{\lambda}_{jk}^{(ScBF)} = \frac{1}{\varsigma_j} \frac{\gamma_{jk}}{d_{jjk}} \tag{46}$$

*where $\varsigma_j$ is given by (20).*

*Proof:* The proof is an easy extension of [12, Theorem 1] derived for a single cell network. ∎

As expected, the computation of $\{\overline{\lambda}_{jk}^{(ScBF)}\}$ for cell $j$ requires only knowledge of the SINR constraints and average channel attenuations of its own cells, i.e., $\{\gamma_{jk}; \forall k\}$ and $\{d_{jjk}; \forall k\}$. Replacing $\lambda_{jk}^\star$ in (45) with $\overline{\lambda}_{jk}^{(ScBF)}$ in (46), we then have the following result.

**Lemma 5.** *Under Assumptions 1, 2 and 4,* $\max_{jk} |\mathrm{SINR}_{jk} - \overline{\mathrm{SINR}}_{jk}^{(ScBF)}| \to 0$ *almost surely with*

$$\overline{\mathrm{SINR}}_{jk}^{(ScBF)} = \left(1 - \frac{1}{N}\sum_{i=1}^{K}\frac{\gamma_{ji}^2}{(1+\gamma_{ji})^2}\right)\frac{p_{jk}d_{jjk}\left(1 - \tau_{jjk}^2\right)}{\overline{I}_{jk}^{(ScBF)} + \sigma^2} \tag{47}$$

*where* $\overline{I}_{jk}^{(ScBF)} \triangleq \sum_{l=1}^{L}\alpha_{ljk}\left(\frac{1}{N}\sum_{i=1}^{K}p_{li}\right)$ *accounts for the interference with*

$$\alpha_{ljk} \triangleq \begin{cases} d_{jjk}\frac{1 - \tau_{jjk}^2\left[1 - (1+\gamma_{jk})^2\right]}{(1+\gamma_{jk})^2} & j = l \\ d_{ljk} & j \neq l. \end{cases} \tag{48}$$

*Proof:* The proof is the same as for Lemma 1, the main difference lying in the fact that $\widehat{v}_{lk}^\star$ in (22) is now independent from $\mathbf{h}_{ljk}$ as shown in Section II-C. ∎

**Theorem 6.** *Let* $\boldsymbol{\Delta} \in \mathbb{C}^{L \times L}$ *be a diagonal matrix with elements*

$$[\boldsymbol{\Delta}]_{j,j} \triangleq 1 - \frac{1}{N}\sum_{k=1}^{K}\frac{\gamma_{jk}^2}{(1+\gamma_{jk})^2} \tag{49}$$

*and* $\mathbf{U} \in \mathbb{C}^{L \times L}$ *such that its $(j,l)$th element is*

$$[\mathbf{U}]_{j,l} \triangleq \frac{1}{N}\sum_{k=1}^{K}\frac{\gamma_{jk}\alpha_{ljk}}{d_{jjk}(1 - \tau_{jjk}^2)} \tag{50}$$

*where* $\alpha_{ljk}$ *is defined in (48). If and only if* $\limsup_K \|\boldsymbol{\Delta}^{-1}\mathbf{U}\| < 1$, *then under Assumptions 1, 2 and 4, the powers* $\{p_{jk}\}$ *required to meet the SINR constraints in the asymptotic regime are:*

$$\overline{p}_{jk}^{(ScBF)} = \frac{\gamma_{jk}}{d_{jjk}\left(1 - \tau_{jjk}^2\right)}\frac{\sum_{l=1}^{L}\alpha_{ljk}\overline{P}_l^{(ScBF)} + \sigma^2}{1 - \frac{1}{N}\sum_{i=1}^{K}\frac{\gamma_{ji}^2}{(1+\gamma_{ji})^2}} \tag{51}$$

*where* $\overline{\mathbf{P}}^{(ScBF)} = [\overline{P}_1^{(ScBF)}, \ldots, \overline{P}_L^{(ScBF)}]^T = \sigma^2(\boldsymbol{\Delta} - \mathbf{U})^{-1}\mathbf{b}$ *collects the total transmit power of each BS. Moreover, the asymptotic total transmit power is*

$$\overline{P}_T^{(ScBF)} = \mathbf{1}^T\overline{\mathbf{P}}^{(ScBF)} = \sigma^2\mathbf{1}^T(\boldsymbol{\Delta} - \mathbf{U})^{-1}\mathbf{b}. \tag{52}$$

*Proof:* The proof is similar to that of Theorem 2. ∎

As seen, the computation of $\{\overline{\lambda}_{jk}^{(ScBF)}\}$ and $\{\overline{p}_{jk}^{(ScBF)}\}$ requires the BSs to exchange almost the same system parameters of CoBF. Then, one might argue that ScBF has no potential advantage with respect to CoBF. The advantage comes from the observation that, unlike CoBF, the implementation of ScBF does not require knowledge of $\{\mathbf{h}_{jli}; \forall l \neq j, i\}$ at each cell $j$. Although it is true that this information could be potentially acquired at cell $j$, operating with a time-division duplex protocol, simply by reception of the pilot signals transmitted by all UEs, from a practical standpoint, this task would require a proper allocation of pilot sequences to avoid the so-called pilot contamination effect. In short, compared to CoBF, ScBF allows to simplify the channel estimation task.

As done for CoBF and CoMP, we use [22, Corollary 3] to state that if $\frac{1}{N}\sum_{l=1}^{L}\sum_{k=1}^{K}\frac{\gamma_{jk}\alpha_{ljk}}{d_{jjk}(1-\tau_{jjk}^2)} \leq 1 - \frac{1}{N}\sum_{k=1}^{K}\frac{\gamma_{jk}^2}{(1+\gamma_{jk})^2}$ $\forall j \in \{1, \ldots, L\}$ or, equivalently,

$$\frac{1}{N}\sum_{k=1}^{K}\frac{\gamma_{jk}}{1 - \tau_{jjk}^2}\left(\tau_{jjk}^2 + \frac{1}{d_{jjk}}\sum_{l=1,l\neq j}^{L}d_{ljk}\right) \leq$$
$$1 - \frac{1}{N}\sum_{k=1}^{K}\frac{\gamma_{jk}}{1+\gamma_{jk}} \quad \forall j \in \{1, \ldots, L\} \tag{53}$$

with the strict inequality holding for at least one $j$, then $\limsup_K \|\boldsymbol{\Delta}^{-1}\mathbf{U}\| < 1$. Imposing $\tau_{jjk} = \tau_j$ $\forall k$ and using (53), from the above condition it follows that (after simple calculus):

**Lemma 6.** *Assume that* $\tau_{jjk} = \tau_j$ $\forall k$. *Then, for any given set of* $\{\gamma_{jk}\}$, $\{d_{jjk}\}$ *and* $\{d_{ljk}\}$, *a feasible power allocation exists in the ScBF if* $\tau_j < \tau_{j,\max}^{(ScBF)}$ *with*

$$\tau_{j,\max}^{(ScBF)} = \left(\frac{1 - \frac{1}{N}\sum_{k=1}^{K}\left(\frac{\gamma_{jk}}{1+\gamma_{jk}} + \frac{\gamma_{jk}}{d_{jjk}}\sum_{l=1,l\neq j}^{L}d_{ljk}\right)}{1 - \frac{1}{N}\sum_{k=1}^{K}\frac{\gamma_{jk}^2}{1+\gamma_{jk}}}\right)^{1/2} \tag{54}$$

*and* $\forall j \in \{1, \ldots, L\}$.

If we assume that $\gamma_{jk} = \gamma_j$ $\forall k$, then from (53) the maximum value of $\gamma_j$ in cell $j$ as a function of the other system parameters is obtained as follows:

**Lemma 7.** *Assume* $\gamma_{jk} = \gamma_j$ $\forall k$. *Then, for any given set of* $\{\tau_{jjk}\}$, $\{d_{jjk}\}$ *and* $\{d_{ljk}\}$, *a feasible asymptotic power allocation exists for ScBF if* $\gamma_j < \gamma_{j,\max}$ *with*

$$\gamma_{j,\max} = \frac{A_j + \frac{K}{N} - 1}{2A_j}\left(1 + \sqrt{1 + 4\frac{A_j}{A_j + \frac{K}{N} - 1}}\right) \tag{55}$$

*and* $A_j = \frac{1}{N}\sum_{k=1}^{K}\frac{\tau_{jjk}^2 + \frac{1}{d_{jjk}}\sum_{l=1,l\neq j}^{L}d_{ljk}}{1 - \tau_{jjk}^2}$.

The asymptotic optimality of the power allocation of Theorem 6 is confirmed as follows:

**Corollary 5.** *Under Assumptions 1, 2 and 4, if perfect knowledge of the channel is available, i.e.,* $\tau_{ljk} = 0$ $\forall l, j, k$, *then the duality gap between (11) and (13) is zero, i.e.,* $\frac{1}{N}\sum_{k=1}^{K}\overline{p}_{jk}^{(ScBF)} = \frac{1}{N}\sum_{k=1}^{K}\overline{\lambda}_{jk}^{(ScBF)}\overline{\sigma}_{jk}^2$.

*Proof:* The proof is sketched in Appendix G. ∎

### D. A simple case study

We now consider a simple case study that allows to easily confirm some of the insights observed above (and to get further ones). In particular, we consider a two-cell network, i.e., $L = 2$, wherein a single UE is active in each cell $K = 1$ with $\gamma_{11} = \gamma_{22} = \gamma$. We assume also for simplicity that the network exhibits some symmetry (as shown in Fig. 1). In particular, we assume that UE in cell 1 is positioned at the same distance from BS 1 and 2 such that $d_{111} = d_{211} = d$ whereas UE in cell 2 is such that $d_{221} = d$ and $d_{121} = \alpha d$ with $d$ being the resulting path loss and $\alpha < 1$. With ScBF, we have that $\lambda_{11}^{(ScBF)} = \lambda_{21}^{(ScBF)} = \frac{1}{\varsigma}\frac{\gamma}{d}$ with $\varsigma = \varsigma_1 = \varsigma_2 = 1 - \frac{1}{N}\frac{\gamma}{1+\gamma}$. In the CoBF case, it turns out that $\lambda_{11}^{(CoBF)} = \frac{1}{\eta_1}\frac{\gamma}{d}$ and $\lambda_{21}^{(CoBF)} = \frac{1}{\eta_2}\frac{\gamma}{d}$ with (from (18))

$$\eta_1 = 1 - \frac{1}{N}\left(\frac{\gamma}{1+\gamma} + \frac{\gamma}{\frac{\eta}{\alpha} + \gamma}\right) = \varsigma - \frac{1}{N}\frac{\gamma}{\frac{\eta}{\alpha}+\gamma} \quad (56)$$

$$\eta_2 = 1 - \frac{1}{N}\left(\frac{\gamma}{1+\gamma} + \frac{\gamma}{\frac{1}{\eta} + \gamma}\right) = \varsigma - \frac{1}{N}\frac{\gamma}{\frac{1}{\eta}+\gamma} \quad (57)$$

where we have defined $\eta = \eta_2/\eta_1$. Since $\alpha < 1$, it easily follows (by contradiction) that $\eta_1$ must be larger than $\eta_2$ such that $\lambda_{11} < \lambda_{21}$. As intuitively expected, higher priority is given to cell 2 since it creates higher interference than cell 1 due to the shorter relative strength. As for the CoMP, it turns out that $\epsilon_1 = d/2\left(\mu_1 + \mu_2\right)$ and $\epsilon_2 = d/2\left(\alpha\mu_1 + \mu_2\right)$ from which one gets $\lambda_1 = \frac{2}{\mu_1 + \mu_2}\frac{\gamma}{d}$ and $\lambda_2 = \frac{2}{\alpha\mu_1 + \mu_2}\frac{\gamma}{d}$. Since $\alpha < 1$, then $\lambda_2 > \lambda_1$. Therefore, higher priority is given to user 2, as it should since BS 1 is located further away from user 2. If a symmetric network is considered such that for example $\alpha = 1$, then it turns out that $\eta_1 = \eta_2 = \eta$ with $\eta = 1 - \frac{2}{N}\frac{\gamma}{1+\gamma}$ and that $\epsilon_1 = \epsilon_2 = d/2\left(\mu_1 + \mu_2\right)$.

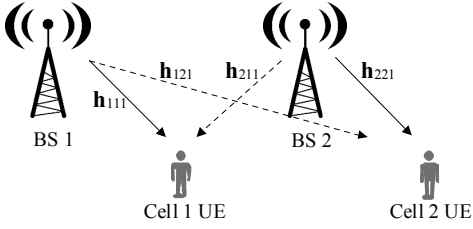

Fig. 1. Graphical representation of a simple two-cell network with one UE per cell such that $d_{111} = d_{211} = d_{221} = d$ and $d_{121} = \alpha d$ with $\alpha < 1$.

From Lemma 6, the maximum value of imperfect CSI in the ScBF configuration is found to be:

$$\tau_{1,\max}^{(ScBF)} = \left(\frac{1 - \frac{1}{N}\left(\frac{\gamma}{1+\gamma} + 1\right)}{1 - \frac{1}{N}\frac{\gamma^2}{1+\gamma}}\right)^{1/2} \quad (58)$$

$$\tau_{2,\max}^{(ScBF)} = \left(\frac{1 - \frac{1}{N}\left(\frac{\gamma}{1+\gamma} + \alpha\right)}{1 - \frac{1}{N}\frac{\gamma^2}{1+\gamma}}\right)^{1/2} \quad (59)$$

from which, it follows that, since $\alpha < 1$, it follows (as it should be due to the higher interference experienced by cell 1) $\tau_{1,\max}^{(ScBF)} < \tau_{2,\max}^{(ScBF)}$. Moreover, from Lemma 2 we obtain

$$\tau_{1,\max}^{(CoBF)} = \left(1 + \frac{\frac{2}{N}\gamma}{1 - \frac{1}{N}\left(\frac{\gamma}{1+\gamma} + \frac{\left(\frac{\gamma}{\eta}\right)^2+\gamma}{\left(1+\frac{\gamma}{\eta}\right)^2}\right)}\right)^{-1/2} \quad (60)$$

$$\tau_{2,\max}^{(CoBF)} = \left(1 + \frac{\frac{1}{N}\gamma\left(1+\alpha\right)}{1 - \frac{1}{N}\left(\frac{\gamma}{1+\gamma} + \frac{1}{N}\frac{\left(\gamma\frac{\eta}{\alpha}\right)^2+\frac{\gamma}{\alpha}}{\left(1+\gamma\frac{\eta}{\alpha}\right)^2}\right)}\right)^{-1/2} \quad (61)$$

with $\tau_{1,\max}^{(CoBF)} < \tau_{2,\max}^{(CoBF)}$ as it follows from standard analysis taking into account that $\eta < 1$ and $\alpha < 1$. Finally, the values of $\tau_{1,\max}^{(CoMP)}$ and $\tau_{2,\max}^{(CoMP)}$ can be obtained from Lemma 4. The lack of explicit expressions for $\{\epsilon_i\}$ and $\{\epsilon'_{i,k}\}$ does not allow an easy comparison between the two values. However, numerical results can be used to confirm the intuition that $\tau_{1,\max}^{(CoMP)} < \tau_{2,\max}^{(CoMP)}$.

*E. On the limiting case $N \to \infty$ and $K/N \to 0$*

We now look at the limiting case in which $N \to \infty$ and $K/N \to 0$. The following results are easily obtained from the asymptotic analysis above:

**Corollary 6.** *If $N \to \infty$ with $K/N \to 0$, then $\widehat{\mathbf{v}}_{jk}^{\star} = \widehat{\mathbf{h}}_{jk}$ and*

$$\overline{p}_{jk}^{(CoBF)} = \frac{\sigma^2}{1 - \tau_{jjk}^2}\frac{\gamma_{jk}}{d_{jjk}}. \quad (62)$$

*Also, we have that $P_T^{(CoBF)} \to 0$ such that*

$$N\overline{P}_T^{(CoBF)} = \sum_{l=1}^{L}\sum_{i=1}^{K}\frac{\sigma^2}{1 - \tau_{lli}^2}\frac{\gamma_{li}}{d_{lli}}. \quad (63)$$

*Proof:* The proof follows from Theorems 1 and 2 observing that if $N \to \infty$ such that $K/N \to 0$, then $\eta_j \to 1$, $\overline{\lambda}_{jk}^{(CoBF)} = \gamma_{jk}/d_{jjk}$, $\sum_{l=1}^{L}\sum_{i=1}^{K}\frac{\overline{\lambda}_{li}^{(CoBF)}}{N}\mathbf{h}_{jli}\mathbf{h}_{jli}^{H} \to 0$. Also, $[\mathbf{\Gamma}]_{j,j} \to 1$ and $[\mathbf{F}]_{j,l} \to 0$ $\forall j, l$. ∎

**Corollary 7.** *If $N \to \infty$ with $K/N \to 0$, then $\widehat{\mathbf{v}}_k^{\star} = \widehat{\mathbf{h}}_k$ and*

$$\overline{p}_k^{(CoMP)} = \frac{\sigma^2}{1 - \tau_k^2}\frac{\gamma_k}{\frac{1}{L}\sum_{l=1}^{L}d_{lk}}. \quad (64)$$

*Also, we have that $\overline{P}_T^{(CoMP)} \to 0$ with*

$$N\overline{P}_T^{(CoMP)} = \sum_{k=1}^{KL}\frac{\sigma^2}{1 - \tau_k^2}\frac{\gamma_k}{\sum_{l=1}^{L}d_{lk}}. \quad (65)$$

*Proof:* The proof follows from Theorems 3 and 4. If $N \to \infty$ and $K/N \to 0$, then $\mu_l \to 1$, $\epsilon'_k \to \frac{1}{L}\sum_{l=1}^{L}d_{li}$, $\overline{\lambda}_k^{(CoMP)} = \gamma_k/(\frac{1}{L}\sum_{l=1}^{L}d_{lk})$ and $\sum_{i=1}^{KL}\frac{\overline{\lambda}_i^{(CoMP)}}{NL}\mathbf{h}_i\mathbf{h}_i^{H} \to 0$. Also, $\overline{\Omega}_k \to 0$ since $z_k \to 0$. ∎

**Corollary 8.** *If $N \to \infty$ with $K/N \to 0$, then ScBF boils down to CoBF.*

The above corollaries state that, if $N$ grows unbounded, the precoder for power minimization reduces to the maximum ratio transmit (MRT) scheme for all network configurations. Moreover, it turns out that the power required by all schemes to meet the constraints $\{\gamma_{li}; \forall l, i\}$ (or, equivalently, $\{\gamma_k; \forall k\}$ for CoMP) vanishes inversely proportional to $1/N$. Moreover, the following result holds true:

**Lemma 8.** *If $L > 2$, $N \to \infty$ and $K$ is kept fixed, then*

$$N\overline{P}_T^{(CoBF)} < N\overline{P}_T^{(CoBF)} = N\overline{P}_T^{(ScBF)} \tag{66}$$

*Proof:* Extending to CoMP the double index notation used for CoBF, (65) becomes $N\overline{P}_T^{(CoMP)} = \sum_{l=1}^{L} \sum_{i=1}^{K} \frac{\sigma^2}{1 - \tau_{lli}^2} \frac{\gamma_{lli}}{\sum_{j=1}^{L} d_{jli}}$ from which the result follows since $\sum_{j=1}^{L} d_{jli} > d_{lli}$ when $L > 2$. ∎

The above result states that the total power consumption decreases faster for CoMP than for CoBF or ScBF, meaning that a fully-cooperative system provides potential advantages for power saving. Also, ScBF performs as CoBF when the number $N$ of antennas grows very large. Consider for example a system in which $\gamma_{lli} = \gamma$ and $\tau_{lli} = \tau$ $\forall l, i$, and, thus, $\gamma_k = \gamma$ and $\tau_k = \tau$ $\forall k$. Assume also that $d_{jli} = d$ $\forall j, l, i$. In these circumstances, as $N$ grows we obtain

$$N\overline{P}_T^{(CoBF)} = N\overline{P}_T^{(ScBF)} = KL\frac{\sigma^2}{1 - \tau^2}\frac{\gamma}{d} \tag{67}$$

$$N\overline{P}_T^{(CoMP)} = K\frac{\sigma^2}{1 - \tau^2}\frac{\gamma}{d} \tag{68}$$

from which it follows that a power saving of $10 \log L$ dB is achieved with CoMP in this particular setting. This can be potentially used to reduce the number of antennas by a factor $L$. The above results are only apparently in contrast to those in [23] wherein it is shown that a CoBF system provides considerable performance improvement (under a wide range of utility functions) as compared to CoMP. Indeed, the results of [23] are obtained under the assumption that the same number of spatial degrees of freedom per UE is provided by each network configuration. This would amount to assuming that each BS in the CoBF and ScBF settings is equipped with $NL$ antennas instead of $N$. Under this assumption, our results corroborate those in [23]. However, this would require to increase the number of antennas per BS by a factor $L$ for CoBF and ScBF.

## IV. Numerical results

Monte-Carlo (MC) simulations are now used to validate the above asymptotic analysis for a network with finite size. The results are obtained for 1000 different channel realizations and UE distributions. We consider a multi-cell network composed of $L$ square cells distributed in a square region of side length 500 m. Following [12], the path loss function $d_{ljk}$ is obtained as $d_{ljk} = 2L_{\overline{x}}(1 + \|\mathbf{x}_{ljk}\|^{\kappa}/\overline{x}^{\kappa})^{-1}$ where $\mathbf{x}_{ljk} \in \mathbb{R}^2$ is the position of user $k$ in cell $j$ with respect to BS $l$, $\kappa > 2$ is the path loss exponent, $\overline{x} > 0$ is some cut-off parameter and $L_{\overline{x}}$ is a constant that regulates the attenuation at distance $\overline{x}$. We assume that $\kappa = 3.5$ and $L_{\overline{x}} = -86.5$ dB [12]. Similarly, we have that $d_{jk} = 2L_{\overline{x}}(1 + \|\mathbf{x}_{jk}\|^{\beta}/\overline{x}^{\beta})^{-1}$ with $\mathbf{x}_{jk}$ being the position of UE $k$ with respect to BS $j$. The transmission bandwidth is $W = 10$ MHz and the total noise power $W\sigma^2$ is $-104$ dBm. Unless otherwise specified, in the subsequent simulations we assume that the same data rate must be guaranteed to each UE. Moreover, we assume that $K = 8$, $N = 32$ and impose the same quality of channel estimate for each UE, i.e., $\tau_{lli} = \tau$ $\forall l, i$.

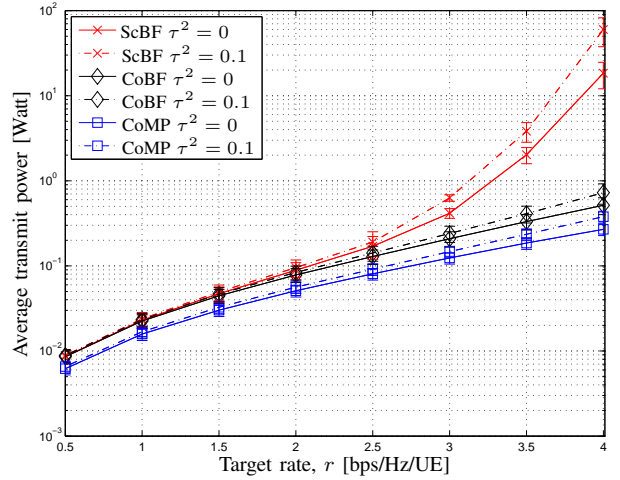

Fig. 2. Average transmit power in Watt vs. target rate when $L = 4$, $K = 8$ and $N = 32$.

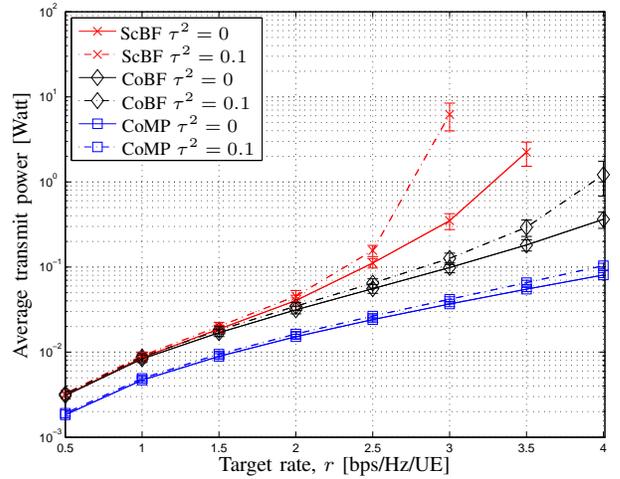

Fig. 3. Average transmit power in Watt vs. target rate when $L = 16$, $K = 8$ and $N = 32$.

Fig. 2 illustrates the average transmit power in Watt vs. target rate in bps/Hz/UE when $L = 4$. Markers are obtained using the asymptotic analysis whereas the error bars indicate the standard deviation of the MC results. Clearly, $\tau^2 = 0$ corresponds to the perfect CSI case. As expected, the higher power consumption is required by ScBF due to the lack of cooperation among BSs. Compared to CoMP, a slight increase of power is required by CoBF. As seen, the approximation lies roughly within one standard deviation of the MC simulations and thus we may conclude that the large system analysis is accurate even for networks of finite size.

We now investigate the performance of the different schemes when the network becomes denser. To this end, Fig. 3 plots the average transmit power in Watt vs. target rate when $L = 16$ such that the total number of UEs in the network is $KL = 128$. Similar conclusions as for Fig. 2 can be drawn with the only difference that the average transmit power for target rates up to 2.5 bps/Hz/UE is smaller for all schemes due to the shorter distances of UEs from their serving BSs. A larger gap between CoBF and CoMP is observed compared to the results of Fig. 2. This means that full cooperation is beneficial as the network becomes denser. This is true especially when imperfect CSI is available and high target rates must be ensured as it can be deducted by the rapid increase of the transmit power for CoBF when $\tau^2 = 0.1$ and $r \geq 3$. Moreover, it can be seen that ScBF cannot support target rates beyond 3.5 and 3 [bps/Hz/UE] for $\tau^2 = 0$ and $\tau^2 = 0.1$, respectively.

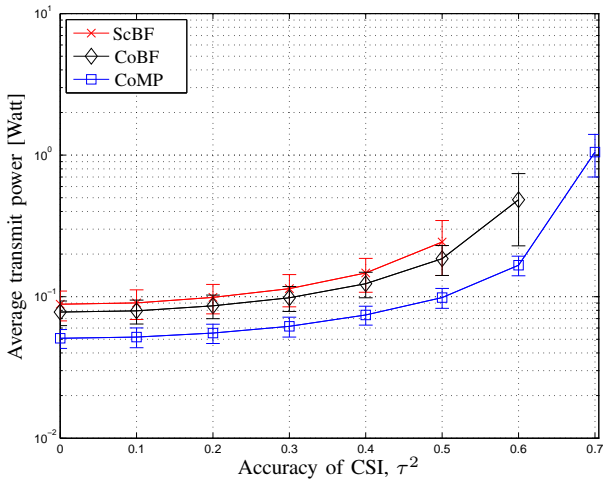

Fig. 4. Average transmit power in Watt vs. $\tau^2$ when $L = 4$, $K = 8$, $N = 32$ and $r = 2$ [bps/Hz/UE].

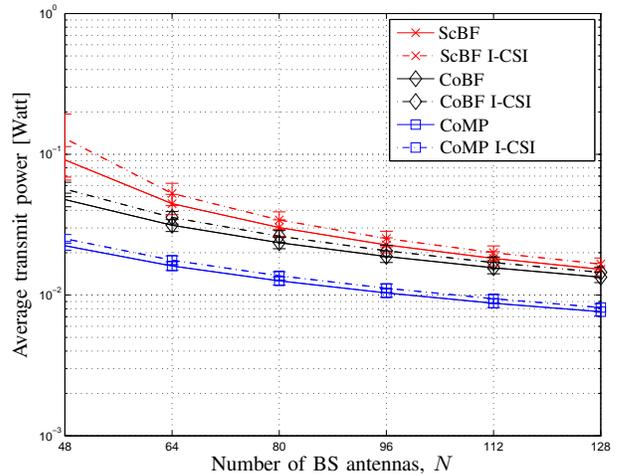

Fig. 6. Average transmit power in Watt vs. $N$ when $L = 16$, $K = 8$ and the rates are randomly taken in the interval $[1, 3]$.

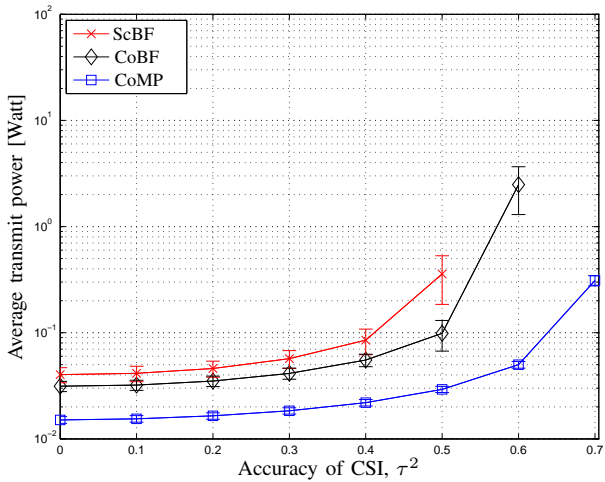

Fig. 5. Average transmit power in Watt vs. $\tau^2$ when $L = 16$, $K = 8$, $N = 32$ and $r = 2$ [bps/Hz/UE].

Figs. 4 and 5 plot the average transmit power in Watt vs. accuracy of CSI when the target rate for all UEs is 2 bps/Hz/UE. As expected, the average transmit power increases as $\tau$ becomes larger and the CoMP technology provides more robustness to imperfect CSI compared to CoBF and ScBF. This holds true also for larger values of target rates.

Fig. 6 plots the average transmit power in Watt vs. $N$ when $K = 8$ and $L = 16$. The label I-CSI refers to the imperfect CSI case. The values of $\{\tau_{lli}^2; \forall l, i\}$ and $\{r_{li}; \forall l, i\}$ are randomly taken in the intervals $[0.01, 0.1]$ and $[1, 3]$ bps/Hz/UE, respectively. As expected, the average transmit power of all schemes decreases as $N$ grows large but, for a given $N$, CoMP requires much less power compared to CoBF and ScBF. The latter perform almost the same as $N$ is large enough.

## V. Conclusions

In this work, we analyzed the structure of the optimal linear precoder for minimizing the total transmit power when different degrees of cooperation among BSs are considered: single cell processing, coordinated beamforming and coordinated multi-point processing. Stating and proving new results from large-scale random matrix theory allowed us to give concise approximations of the Lagrange multipliers, the powers needed to ensure target rates and the total transmit power. Such approximations turned out to depend only on the long-term channel attenuations of the UEs, the relative strength of interference among BSs, the target rates and the quality of the channel estimates. Numerical results indicated that these approximations are very accurate even for small system dimensions. Applied to practical networks, such results may lead to important insights into the system behavior, especially with respect to target rates, CSI quality and induced interference. Moreover, they can be used to simulate the network behavior without to carry out extensive Monte-Carlo simulations. It is worth observing that the asymptotic analysis provided in this article could potentially be extended to more advanced channel models to account for example for the correlation among BS antennas [11] and/or for the presence of line of sight components [24], [25]. All this is left for future work.

## Appendix A
### Proof of Theorem 1

Let us first provide some intuition on the main result. To this end, apply the matrix inversion lemma [26] to rewrite (15) as follows

$$\frac{\gamma_{jk}}{\lambda_{jk}^\star} = \frac{1}{N} \widehat{\mathbf{h}}_{jjk}^H \left( \frac{1}{N} \sum_{(l,i) \neq (j,k)} \lambda_{li}^\star \widehat{\mathbf{h}}_{jli} \widehat{\mathbf{h}}_{jli}^H + \mathbf{I}_N \right)^{-1} \widehat{\mathbf{h}}_{jjk}. \quad (69)$$

Assume erroneously for a moment that the scalars $\{\lambda_{li}^{\star}\}$ are given and independent from the channel vectors $\{\widehat{\mathbf{h}}_{jli}\}$. Then, using classical random matrix theory results such as those in [11, Lemmas 4 and 5] yields

$$\frac{\gamma_{jk}}{\lambda_{jk}^{\star}} \asymp \frac{d_{jjk}}{N} \operatorname{tr}\left(\frac{1}{N} \sum_{l=1}^{L} \sum_{i=1}^{K} \lambda_{li}^{\star} \widehat{\mathbf{h}}_{jli} \widehat{\mathbf{h}}_{jli}^{H} + \mathbf{I}_{N}\right)^{-1}. \quad (70)$$

Using [11, Theorem 1] we have that

$$\frac{1}{N} \operatorname{tr}\left(\frac{1}{N} \sum_{l=1}^{L} \sum_{i=1}^{K} \lambda_{li}^{\star} \widehat{\mathbf{h}}_{jli} \widehat{\mathbf{h}}_{jli}^{H} + \mathbf{I}_{N}\right)^{-1} \asymp \eta_{j}^{\star} \quad (71)$$

where the coefficients $\{\eta_{j}^{\star}\}$ are solutions of the following system of equations

$$\eta_{j}^{\star} = \left(\frac{1}{N} \sum_{l=1}^{L} \sum_{i=1}^{K} \frac{\lambda_{li}^{\star} d_{jli}}{1 + \lambda_{li}^{\star} d_{jli} \eta_{j}^{\star}} + 1\right)^{-1}. \quad (72)$$

From the above discussion, we may then expect the terms $\lambda_{jk}^{\star}$ to be all close to $\lambda_{jk}^{\star} = \frac{1}{\eta_{j}^{\star}} \frac{\gamma_{jk}}{d_{jjk}}$ for $N, K$ large enough. This statement is made rigorous in the following. To this end, let us define

$$c_{jk} = \frac{\overline{\lambda}_{jk}^{(CoBF)}}{\lambda_{jk}^{\star}} = \frac{\gamma_{jk}}{d_{jjk} \eta_{j}} \frac{1}{\lambda_{jk}^{\star}} \quad (73)$$

where the coefficients $\{\eta_{j}\}$ are such that:

$$\eta_{j} = \left(\frac{1}{N} \sum_{l=1}^{L} \sum_{i=1}^{K} \frac{\overline{\lambda}_{li}^{(CoBF)} d_{jli}}{1 + \overline{\lambda}_{li}^{(CoBF)} d_{jli} \eta_{j}} + 1\right)^{-1}. \quad (74)$$

From (69), using (73) one gets

$$\frac{\gamma_{jk} c_{jk}}{\overline{\lambda}_{jk} d_{jjk}} = \frac{1}{N} \mathbf{z}_{jjk}^{H} \left(\frac{1}{N} \sum_{(l,i) \neq (j,k)} \frac{\overline{\lambda}_{li}}{c_{li}} d_{jli} \mathbf{z}_{jli} \mathbf{z}_{jli}^{H} + \mathbf{I}_{N}\right)^{-1} \mathbf{z}_{jjk} \quad (75)$$

where the superscript $^{(CoMP)}$ is omitted for simplicity. Assume that $\{c_{jk}\}$ are well defined, positive and such that $0 \leq c_{11} \leq c_{12} \leq \cdots \leq c_{1K} \leq c_{21} \leq c_{22} \leq \cdots \leq c_{2K} \leq \cdots \leq c_{L1} \leq c_{L2} \leq \cdots \leq c_{LK}$. Then, using monotonicity arguments, from (75) it follows that for $j = L$ and $k = K$ we have

$$\frac{\gamma_{LK}}{\overline{\lambda}_{LK}} \frac{1}{d_{LLK}} \leq \frac{1}{N} \mathbf{z}_{LLK}^{H} \mathbf{B}_{k}^{-1}(c_{LK}) \mathbf{z}_{LLK}. \quad (76)$$

with $\mathbf{B}_{k}(c_{LK}) = \frac{1}{N} \sum_{(l,i) \neq (L,K)} \overline{\lambda}_{li} d_{jli} \mathbf{z}_{jli} \mathbf{z}_{jli}^{H} + c_{LK} \mathbf{I}_{N}$. Assume now that $c_{LK}$ is infinitely often larger than $1 + \ell$ with $\ell > 0$ some positive value [17]. Let us restrict ourselves to such a subsequence. From (76), using monotonicity arguments we obtain

$$\frac{\gamma_{LK}}{\overline{\lambda}_{LK} d_{LLK}} \leq \frac{1}{N} \mathbf{z}_{LLK}^{H} \mathbf{B}_{k}^{-1}(1 + \ell) \mathbf{z}_{LLK}. \quad (77)$$

Applying [11, Theorem 1] one gets

$$\frac{1}{N} \mathbf{z}_{LLK}^{H} \mathbf{B}_{k}^{-1}(1 + \ell) \mathbf{z}_{LLK} \asymp e_{L}(\ell) \quad (78)$$

with $e_{j}(\ell)$ being the unique positive solution to

$$e_{j}(\ell) = \left(\frac{1}{N} \sum_{l=1}^{L} \sum_{i=1}^{K} \frac{\overline{\lambda}_{li} d_{jli}}{1 + \overline{\lambda}_{li} d_{jli} e_{j}(\ell)} + 1 + \ell\right)^{-1}. \quad (79)$$

From (78), recalling (77) yields $\lim_{K \to \infty} \inf e_{L}(\ell) \geq \frac{\gamma_{LK}}{\overline{\lambda}_{LK}} \frac{1}{d_{LLK}}$. Using the fact that $e_{L}(0) = \eta_{L}$ and that $e_{L}(\ell)$ is a decreasing function of $\ell$, it can be proved [17] that for any $\ell > 0$ $\lim_{K \to \infty} \sup e_{L}(\ell) < \frac{\gamma_{LK}}{\overline{\lambda}_{LK}} \frac{1}{d_{LLK}}$. This however goes against the former condition and creates a contradiction on the initial hypothesis that $c_{LK} > 1 + \ell$ infinitely often. Therefore, we must admit that $c_{LK} \leq 1 + \ell$ for all large values of $K$. Reverting all inequalities and using similar arguments yields $c_{L1} \geq 1 - \ell$ for all large values of $K$. Putting all these results together yields $1 - \ell \leq c_{L1} \leq c_{L2} \leq \cdots \leq c_{LK} \leq 1 + \ell$ from which we may write $\max_{k=1,2,\ldots,K} |c_{Lk} - 1| \leq \ell$ for all large values of $K$ [17]. Taking a countable sequence of $\ell$ going to zero, we eventually obtain $\max_{k=1,2,\ldots,K} |c_{Lk} - 1| \to 0$ from which using (73) and assuming that $\lim_{K \to \infty} \sup \frac{\gamma_{LK}}{d_{LLK}} < \infty$ it follows that $\max_{k=1,2,\ldots,K} |\lambda_{Lk}^{\star} - \overline{\lambda}_{Lk}^{(CoBF)}| \to 0$ with $\overline{\lambda}_{Lk}^{(CoBF)} = \frac{1}{\eta_{L}} \frac{\gamma_{Lk}}{d_{LLk}}$. Following the same steps for $j = 1, 2, \ldots, L - 1$ completes the proof.

## Appendix B
## Proof of Lemma 1

To begin with, rewrite the numerator of the SINR in (22) as follows

$$\frac{p_{jk}}{N} \frac{\left|\mathbf{h}_{jjk}^{H} \widehat{\mathbf{v}}_{jk}^{\star}\right|^{2}}{\|\widehat{\mathbf{v}}_{jk}^{\star}\|^{2}} = p_{jk} \frac{\left|\frac{1}{N} \mathbf{h}_{jjk}^{H} \widehat{\mathbf{A}}_{j} \widehat{\mathbf{h}}_{jjk}\right|^{2}}{\frac{1}{N} \mathbf{h}_{jjk}^{H} \widehat{\mathbf{A}}_{j}^{2} \widehat{\mathbf{h}}_{jjk}} \quad (80)$$

with $\widehat{\mathbf{A}}_{j} = (\frac{1}{N} \sum_{\ell=1}^{L} \sum_{i=1}^{K} \overline{\lambda}_{li} \widehat{\mathbf{h}}_{jli} \widehat{\mathbf{h}}_{jli}^{H} + \mathbf{I}_{N})^{-1}$ where we have replaced $\overline{\lambda}_{ji}^{(CoBF)} = \overline{\lambda}_{ji}$ for notational simplicity. Applying the matrix inversion lemma first [26] and then using (14) we have that the numerator in the right-hand-side of the above equation takes the form:

$$\frac{1}{N} \mathbf{h}_{jjk}^{H} \widehat{\mathbf{A}}_{j} \widehat{\mathbf{h}}_{jjk} = \frac{\frac{1}{N} \sqrt{1 - \tau_{jk}^{2}} \mathbf{h}_{jjk}^{H} \widehat{\mathbf{A}}_{j}^{[jk]} \mathbf{h}_{jjk}}{1 + \frac{1}{N} \overline{\lambda}_{jk} \widehat{\mathbf{h}}_{jjk}^{H} \widehat{\mathbf{A}}_{j}^{[jk]} \widehat{\mathbf{h}}_{jjk}}$$
$$+ \frac{\frac{1}{N} \tau_{jk} \mathbf{h}_{jjk}^{H} \widehat{\mathbf{A}}_{j}^{[jk]} \mathbf{q}_{jjk}}{1 + \frac{1}{N} \overline{\lambda}_{jk} \widehat{\mathbf{h}}_{jjk}^{H} \widehat{\mathbf{A}}_{j}^{[jk]} \widehat{\mathbf{h}}_{jjk}} \quad (81)$$

where $\widehat{\mathbf{A}}_{j} = \left(\frac{1}{N} \sum_{\ell=1}^{L} \sum_{i=1}^{K} \overline{\lambda}_{li} \widehat{\mathbf{h}}_{jli} \widehat{\mathbf{h}}_{jli}^{H} + \mathbf{I}_{N}\right)^{-1}$ and

$$\widehat{\mathbf{A}}_{j}^{[jk]} = \left(\frac{1}{N} \sum_{(l,i) \neq (j,k)} \overline{\lambda}_{li} \widehat{\mathbf{h}}_{jli} \widehat{\mathbf{h}}_{jli}^{H} + \mathbf{I}_{N}\right)^{-1}. \quad (82)$$

According to [11, Lemmas 4 and 5] and the results of Appendix A as well as Theorem 1 it follows that $\frac{1}{N} \mathbf{h}_{jjk}^{H} \widehat{\mathbf{A}}_{j}^{[jk]} \mathbf{h}_{jjk} \asymp d_{jjk} \eta_{j}$, $\frac{1}{N} \overline{\lambda}_{jk} \widehat{\mathbf{h}}_{jjk}^{H} \widehat{\mathbf{A}}_{j}^{[jk]} \widehat{\mathbf{h}}_{jjk} \asymp \gamma_{j}$ and $\frac{1}{N} \mathbf{h}_{jjk}^{H} \widehat{\mathbf{A}}_{j}^{[jk]} \mathbf{q}_{jjk} \asymp 0$. Therefore, we have that

$$\frac{1}{N} \mathbf{h}_{jjk}^{H} \widehat{\mathbf{A}}_{j} \widehat{\mathbf{h}}_{jjk} \asymp \sqrt{1 - \tau_{jk}^{2}} \frac{d_{jjk} \eta_{j}}{1 + \gamma_{jk}}. \quad (83)$$

We proceed to computing the deterministic approximation of $\frac{1}{N}\widehat{\mathbf{h}}_{jjk}^H \widehat{\mathbf{A}}_j^2 \widehat{\mathbf{h}}_{jjk}$ in the denominator of the right-hand-side of (80). Observe that applying the matrix inversion lemma [26] and using [11, Lemma 4] together with the results of Theorem 1 (such that $\gamma_{jk} = \lambda_{jk} d_{jjk} \eta_j$) one obtain

$$\frac{1}{N}\mathbf{h}_{jjk}^H \widehat{\mathbf{A}}_j^2 \mathbf{h}_{jjk} \asymp \frac{d_{jjk}\mathrm{tr}(\widehat{\mathbf{A}}_j^2)}{(1+\gamma_{jk})^2} \quad (84)$$

with $\frac{1}{N}\mathrm{tr}(\widehat{\mathbf{A}}_j^2) = \frac{1}{N}\frac{\partial}{\partial z}\mathrm{tr}(\widehat{\mathbf{A}}_j^{-1} - z\mathbf{I}_N)^{-1}|_{z=0}$ [2, Theorem 4]. Using similar arguments as those of Appendix A, we have that

$$\frac{1}{N}\mathrm{tr}\left(\widehat{\mathbf{A}}_j^{-1} - z\mathbf{I}_N\right)^{-1} \asymp \eta_j(z) \quad (85)$$

with $\eta_j(z) = \left(\frac{1}{N}\sum_{l=1}^L \sum_{i=1}^K \frac{\gamma_{li}\frac{d_{jli}}{d_{lli}}\frac{1}{\eta_l(z)}}{1+\gamma_{li}\frac{d_{jli}}{d_{lli}}\frac{\eta_j(z)}{\eta_l(z)}} + 1 - z\right)^{-1}$. By differentiating along $z$, we get

$$\eta_j'(z) = \eta_j^2(z)\left(\frac{1}{N}\sum_{l=1}^L \sum_{i=1}^K \frac{\left(\gamma_{li}\frac{d_{jli}}{d_{lli}}\right)^2\frac{\eta_j'(z)}{\eta_l^2(z)}}{\left(1+\gamma_{li}\frac{d_{jli}}{d_{lli}}\frac{\eta_j(z)}{\eta_l(z)}\right)^2} + 1\right) \quad (86)$$

from which setting $z = 0$ and using simple calculus one obtain (omitting the functional dependence from $z = 0$)

$$\eta_j' \triangleq \frac{\eta_j^2}{1 - \frac{1}{N}\sum_{m=1}^L \sum_{i=1}^K \frac{\left(\gamma_{li}\frac{d_{jli}}{d_{lli}}\frac{\eta_j}{\eta_l}\right)^2}{\left(1+\gamma_{li}\frac{d_{jli}}{d_{lli}}\frac{\eta_j}{\eta_l}\right)^2}}. \quad (87)$$

Therefore, $\frac{1}{N}\mathrm{tr}(\widehat{\mathbf{A}}_j^2) \asymp \eta_j'$ such that $\frac{1}{N}\widehat{\mathbf{h}}_{jjk}^H \widehat{\mathbf{A}}_j^2 \widehat{\mathbf{h}}_{jjk} \asymp \frac{d_{jjk}\eta_j'}{(1+\gamma_{jk})^2}$. Putting this result together with (83) we have that:

$$\frac{p_{jk}}{N}\frac{|\mathbf{h}_{jjk}^H \widehat{\mathbf{v}}_{jk}^*|^2}{||\widehat{\mathbf{v}}_{jk}^*||^2} \asymp$$
$$p_{jk}d_{jjk}\left(1 - \tau_{jjk}^2\right)\left(1 - \frac{1}{N}\sum_{l=1}^L \sum_{i=1}^K \frac{\left(\gamma_{li}\frac{d_{jli}}{d_{lli}}\frac{\eta_j}{\eta_l}\right)^2}{\left(1+\gamma_{li}\frac{d_{jli}}{d_{lli}}\frac{\eta_j}{\eta_l}\right)^2}\right).$$

We now deal with the intracell interference term in the denominator of (22), which can be rewritten as

$$\sum_{i=1,i\neq k}^K \frac{p_{ji}}{N}\frac{|\mathbf{h}_{jjk}^H \widehat{\mathbf{v}}_{ji}^*|^2}{||\widehat{\mathbf{v}}_{ji}^*||^2} =$$
$$\frac{1}{N}\mathbf{h}_{jjk}^H \widehat{\mathbf{A}}_j\left(\frac{1}{N}\widehat{\mathbf{H}}_j^{[jk]}\mathbf{P}_j^{[jk]}\widehat{\mathbf{H}}_j^{[jk]^H}\right)\widehat{\mathbf{A}}_j\mathbf{h}_{jjk} \quad (88)$$

with $\widehat{\mathbf{H}}_j^{[jk]} \triangleq [\widehat{\mathbf{h}}_{jj1},\ldots,\widehat{\mathbf{h}}_{jj(k-1)},\widehat{\mathbf{h}}_{jj(k+1)},\ldots\widehat{\mathbf{h}}_{jjK}] \in \mathbb{C}^{N\times K-1}$ and $\mathbf{P}_j^{[jk]} \triangleq \mathrm{diag}\{\frac{p_{j1}}{\frac{1}{N}||\widehat{\mathbf{v}}_{j1}^*||^2},\ldots,\frac{p_{j(k-1)}}{\frac{1}{N}||\widehat{\mathbf{v}}_{j(k-1)}^*||^2},\frac{p_{j(k+1)}}{\frac{1}{N}||\widehat{\mathbf{v}}_{j(k+1)}^*||^2},\ldots,\frac{p_{jK}}{\frac{1}{N}||\widehat{\mathbf{v}}_{jK}^*||^2}\}$. In order to eliminate the dependence between $\mathbf{h}_{jjk}$ and $\widehat{\mathbf{A}}_j$, rewrite (88) as

$$\frac{1}{N}\mathbf{h}_{jjk}^H \widehat{\mathbf{A}}_j\left(\frac{1}{N}\widehat{\mathbf{H}}_j^{[jk]}\mathbf{P}_j^{[jk]}\widehat{\mathbf{H}}_j^{[jk]^H}\right)\widehat{\mathbf{A}}_j\mathbf{h}_{jjk}$$
$$= \frac{1}{N}\mathbf{h}_{jjk}^H \widehat{\mathbf{A}}_j^{[jk]}\left(\frac{1}{N}\widehat{\mathbf{H}}_j^{[jk]}\mathbf{P}_j^{[jk]}\widehat{\mathbf{H}}_j^{[jk]^H}\right)\widehat{\mathbf{A}}_j\mathbf{h}_{jjk} +$$
$$+ \frac{1}{N}\mathbf{h}_{jjk}^H\left(\widehat{\mathbf{A}}_j - \widehat{\mathbf{A}}_j^{[jk]}\right)\left(\frac{1}{N}\widehat{\mathbf{H}}_j^{[jk]}\mathbf{P}_j^{[jk]}\widehat{\mathbf{H}}_j^{[jk]^H}\right)\widehat{\mathbf{A}}_j\mathbf{h}_{jjk}. \quad (89)$$

Using the resolvent identity $\widehat{\mathbf{A}}_j - \widehat{\mathbf{A}}_j^{[jk]} = -\widehat{\mathbf{A}}_j(\widehat{\mathbf{A}}_j^{-1} - \widehat{\mathbf{A}}_j^{[k]^{-1}})\widehat{\mathbf{A}}_j^{[jk]}$ [26] and observing that

$$\widehat{\mathbf{A}}_j^{-1} - \widehat{\mathbf{A}}_j^{[k]^{-1}} = \frac{\overline{\lambda}_{jk}}{N}\Big(c_0\mathbf{h}_{jjk}\mathbf{h}_{jjk}^H + c_1\mathbf{q}_{jjk}\mathbf{q}_{jjk}^H$$
$$+ c_2\mathbf{h}_{jjk}\mathbf{q}_{jjk}^H + c_2\mathbf{q}_{jjk}\mathbf{h}_{jjk}^H\Big) \quad (90)$$

with $c_0 = 1 - \tau_{jjk}^2, c_1 = \tau_{jjk}^2$ and $c_2 = \tau_{jjk}\sqrt{1 - \tau_{jjk}^2}$, from (89) one gets

$$\frac{1}{N}\mathbf{h}_{jjk}^H \widehat{\mathbf{A}}_j\left(\frac{1}{N}\widehat{\mathbf{H}}_j^{[jk]}\mathbf{P}_j^{[jk]}\widehat{\mathbf{H}}_j^{[jk]^H}\right)\widehat{\mathbf{A}}_j\mathbf{h}_{jjk}$$
$$= \frac{1}{N}\mathbf{h}_{jjk}^H \widehat{\mathbf{B}}_j\mathbf{h}_{jjk} - \overline{\lambda}_{jk}c_0\frac{1}{N}\mathbf{h}_{jjk}^H \widehat{\mathbf{A}}_j\mathbf{h}_{jjk}\frac{1}{N}\mathbf{h}_{jjk}^H \widehat{\mathbf{B}}_j\mathbf{h}_{jjk} -$$
$$- \overline{\lambda}_{jk}c_2\frac{1}{N}\mathbf{h}_{jjk}^H \widehat{\mathbf{A}}_j\mathbf{h}_{jjk}\frac{1}{N}\mathbf{q}_{jjk}^H \widehat{\mathbf{B}}_j\mathbf{h}_{jjk}$$
$$- \overline{\lambda}_{jk}c_2\frac{1}{N}\mathbf{h}_{jjk}^H \widehat{\mathbf{A}}_j\mathbf{q}_{jjk}\frac{1}{N}\mathbf{h}_{jjk}^H \widehat{\mathbf{B}}_j\mathbf{h}_{jjk} \quad (91)$$

where we have defined $\widehat{\mathbf{B}}_j = \widehat{\mathbf{A}}_j^{[jk]}(\frac{1}{N}\widehat{\mathbf{H}}_j^{[jk]}\mathbf{P}_j^{[jk]}\widehat{\mathbf{H}}_j^{[jk]^H})\widehat{\mathbf{A}}_j$. Using [11, Lemma 7] we obtain that $\frac{1}{N}\mathbf{h}_{jjk}^H \widehat{\mathbf{B}}_j\mathbf{h}_{jjk} \asymp \frac{u'(1+\overline{\lambda}_{jk}c_1 u)}{1+\overline{\lambda}_{jk}u}$, $\frac{1}{N}\mathbf{h}_{jjk}^H \widehat{\mathbf{A}}_j\mathbf{h}_{jjk} \asymp \frac{u(1+\overline{\lambda}_{jk}c_1 u)}{1+\overline{\lambda}_{jk}u}$, $\frac{1}{N}\mathbf{h}_{jjk}^H \widehat{\mathbf{A}}_j\mathbf{q}_{jjk} \asymp \frac{-\overline{\lambda}_{jk}c_2 u^2}{1+\overline{\lambda}_{jk}u}$ and $\frac{1}{N}\mathbf{q}_{jjk}^H \widehat{\mathbf{B}}_j\mathbf{h}_{jjk} \asymp \frac{-\overline{\lambda}_{jk}c_2 uu'}{1+\overline{\lambda}_{jk}u}$ where we have defined, for notational convenience,

$$u = \frac{d_{jjk}}{N}\mathrm{tr}\left(\widehat{\mathbf{A}}_j^{[jk]}\right) + o(1) \quad (92)$$
$$u' = \frac{d_{jjk}}{N}\mathrm{tr}\left(\frac{1}{N}\mathbf{P}_j^{[jk]}\widehat{\mathbf{H}}_j^{[jk]^H}\widehat{\mathbf{A}}_j^{[jk]^2}\widehat{\mathbf{H}}_j^{[jk]^H}\right) + o(1). \quad (93)$$

Using the above results, (91) can be approximated as

$$\frac{1}{N}\mathbf{h}_{jjk}^H \widehat{\mathbf{A}}_j\left(\frac{1}{N}\widehat{\mathbf{H}}_j^{[jk]}\mathbf{P}_j^{[jk]}\widehat{\mathbf{H}}_j^{[jk]^H}\right)\widehat{\mathbf{A}}_j\mathbf{h}_{jjk} \asymp$$
$$\frac{u'(1+\overline{\lambda}_{jk}\tau_{jjk}^2 u)}{1+\overline{\lambda}_{jk}u} - \overline{\lambda}_{jk}\frac{1-\tau_{jjk}^2}{\left(1+\overline{\lambda}_{jk}u\right)^2}uu' \quad (94)$$

or, equivalently, after simple calculus

$$\frac{1}{N}\mathbf{h}_{jjk}^H \widehat{\mathbf{A}}_j\left(\frac{1}{N}\widehat{\mathbf{H}}_j^{[jk]}\mathbf{P}_j^{[jk]}\widehat{\mathbf{H}}_j^{[jk]^H}\right)\widehat{\mathbf{A}}_j\mathbf{h}_{jjk} \asymp$$
$$\frac{1-\tau_{jjk}^2\left[1-\left(1+\overline{\lambda}_{jk}u\right)^2\right]}{\left(1+\overline{\lambda}_{jk}u\right)^2}u'. \quad (95)$$

Notice that $u$ in (92) is such that $u \asymp \frac{d_{jjk}}{N}\eta_j$ (as it follows from Theorem 1) while the deterministic approximation of $u'$ is computed as follows. Observe that

$$\frac{1}{N}\mathrm{tr}\left(\frac{1}{N}\mathbf{P}_j^{[jk]}\widehat{\mathbf{H}}_j^{[jk]^H}\widehat{\mathbf{A}}_j^{[jk]^2}\widehat{\mathbf{H}}_j^{[jk]}\right) =$$
$$\frac{1}{N}\sum_{i=1,i\neq j}^K p_{ji}\frac{\frac{1}{N}\mathbf{h}_{jji}^H \widehat{\mathbf{A}}_j^{[jk]^2}\widehat{\mathbf{h}}_{jji}}{\frac{1}{N}\widehat{\mathbf{h}}_{jjk}^H \widehat{\mathbf{A}}_j^2\widehat{\mathbf{h}}_{jjk}} \quad (96)$$

from which using [11, Lemmas 4, 5 and 6] one gets

$$\frac{1}{N}\sum_{i=1,i\neq j}^K p_{ji}\frac{\frac{1}{N}\mathbf{h}_{jji}^H \widehat{\mathbf{A}}_j^{[jk]^2}\widehat{\mathbf{h}}_{jji}}{\frac{1}{N}\widehat{\mathbf{h}}_{jjk}^H \widehat{\mathbf{A}}_j^2\widehat{\mathbf{h}}_{jjk}} \asymp \frac{1}{N}\sum_{i=1}^K p_{ji}. \quad (97)$$

Plugging the above result into (93) produces $u' \asymp d_{jjk}(\frac{1}{N}\sum_{i=1}^{K} p_{ji})$. Therefore, recalling (46) we have that

$$\frac{1}{N}\mathbf{h}_{jjk}^{H}\widehat{\mathbf{A}}_{j}\left(\frac{1}{N}\widehat{\mathbf{H}}_{j}^{[jk]}\mathbf{P}_{j}^{[jk]}\widehat{\mathbf{H}}_{j}^{[jk]^{H}}\right)\widehat{\mathbf{A}}_{j}\mathbf{h}_{jjk} \asymp$$

$$d_{jjk}\frac{1-\tau_{jjk}^{2}\left[1-(1+\gamma_{jk})^{2}\right]}{(1+\gamma_{jk})^{2}}\left(\frac{1}{N}\sum_{i=1}^{K}p_{ji}\right). \quad (98)$$

We are left with computing the deterministic equivalent of the intercell interference generated to user $k$ in cell $j$ by all other UEs in cell $l$ $\forall l \neq j$:

$$\sum_{i=1}^{K}\frac{p_{li}}{N}\frac{|\mathbf{h}_{lj k}^{H}\widehat{\mathbf{v}}_{li}^{*}|^{2}}{\|\widehat{\mathbf{v}}_{li}^{*}\|^{2}} = \frac{1}{N}\mathbf{h}_{ljk}^{H}\widehat{\mathbf{A}}_{l}\left(\frac{1}{N}\widehat{\mathbf{H}}_{l}\mathbf{P}_{l}\widehat{\mathbf{H}}_{l}^{H}\right)\widehat{\mathbf{A}}_{l}\mathbf{h}_{ljk} \quad (99)$$

with $\widehat{\mathbf{H}}_{l} \triangleq [\widehat{\mathbf{h}}_{ll1},\ldots,\widehat{\mathbf{h}}_{llK}] \in \mathbb{C}^{N \times K}$ and $\mathbf{P}_{l} \triangleq \text{diag}\{\frac{p_{l1}}{N\|\widehat{\mathbf{v}}_{l1}^{*}\|^{2}},\ldots,\frac{p_{lK}}{N\|\widehat{\mathbf{v}}_{lK}^{*}\|^{2}}\}$. Mimicking the same steps of above one gets

$$\frac{1}{N}\mathbf{h}_{ljk}^{H}\widehat{\mathbf{A}}_{l}\left(\frac{1}{N}\widehat{\mathbf{H}}_{l}\mathbf{P}_{l}\widehat{\mathbf{H}}_{l}^{H}\right)\widehat{\mathbf{A}}_{l}\mathbf{h}_{ljk} \asymp$$

$$\frac{1-\tau_{ljk}^{2}\left[1-(1+\overline{\lambda}_{jk}u)^{2}\right]}{(1+\overline{\lambda}_{jk}u)^{2}}u' \quad (100)$$

where we have that $u = \frac{d_{ljk}}{N}\text{tr}(\widehat{\mathbf{A}}_{l}^{[jk]}) + o(1)$ and $u' = \frac{d_{ljk}}{N}\text{tr}(\frac{1}{N}\mathbf{P}_{l}\widehat{\mathbf{H}}_{l}^{H}\widehat{\mathbf{A}}_{l}^{[jk]^{2}}\widehat{\mathbf{H}}_{l}) + o(1)$. Observe now that $u \asymp d_{ljk}\eta$ and $u' \asymp \frac{d_{ljk}}{N}\sum_{i=1}^{K}p_{li}$. Putting these results together and recalling (16) we eventually obtain

$$\frac{1}{N}\mathbf{h}_{ljk}^{H}\widehat{\mathbf{A}}_{l}\left(\frac{1}{N}\widehat{\mathbf{H}}_{l}\mathbf{P}_{l}\widehat{\mathbf{H}}_{l}^{H}\right)\widehat{\mathbf{A}}_{l}\mathbf{h}_{ljk} \asymp$$

$$d_{ljk}\frac{1-\tau_{ljk}^{2}\left[1-\left(1+\gamma_{jk}\frac{d_{ljk}}{d_{jjk}}\frac{\eta_{l}}{\eta_{j}}\right)^{2}\right]}{\left(1+\gamma_{jk}\frac{d_{ljk}}{d_{jjk}}\frac{\eta_{l}}{\eta_{j}}\right)^{2}}\left(\frac{1}{N}\sum_{i=1}^{K}p_{li}\right). \quad (101)$$

Putting all the above results together completes the proof.

## APPENDIX C
## PROOF OF THEOREM 2

From (23), omitting the superscript $^{(CoBF)}$ it follows that the power $\overline{p}_{jk}$ such that $\overline{\text{SINR}}_{jk} = \gamma_{jk}$ is obtained as:

$$\overline{p}_{jk} = \frac{\eta_{j}'}{\eta_{j}^{2}}\frac{\gamma_{jk}}{d_{jjk}\left(1-\tau_{jjk}^{2}\right)}\left(\overline{I}_{jk}+\sigma^{2}\right) \quad (102)$$

with $\eta_{j}'$ given by (87). Plugging the above result into $\overline{P}_{j} = \frac{1}{N}\sum_{k=1}^{K}\overline{p}_{jk}$ and observing that $\overline{I}_{jk} = \sum_{l=1}^{L}\beta_{ljk}\overline{P}_{l}$ it follows that the values $\{\overline{P}_{j}\}$ must satisfy the following set of equations $\forall j$:

$$\overline{P}_{j}\frac{\eta_{j}^{2}}{\eta_{j}'} = \frac{1}{N}\sum_{k=1}^{K}\frac{\gamma_{jk}}{d_{jjk}(1-\tau_{jjk}^{2})}\left(\sum_{l=1}^{L}\beta_{ljk}\overline{P}_{l}+\sigma^{2}\right)$$

$$= \sum_{l=1}^{L}\overline{P}_{l}[\mathbf{F}]_{j,l}+\sigma^{2}b_{j} \quad (103)$$

where $b_{j}$ and $[\mathbf{F}]_{j,l}$ are given by (25) and (27). Rewriting the above set of equations in matrix form yields $(\mathbf{\Gamma}-\mathbf{F})\overline{\mathbf{P}} = \sigma^{2}\mathbf{b}$ with $\mathbf{\Gamma}$ diagonal and given by (26). Using [22, Theorem 2.1], a necessary and sufficient condition for a solution $\overline{\mathbf{P}} \geq 0$ to exist is that the spectral radius of $\mathbf{\Gamma}^{-1}\mathbf{F}$ is smaller than 1. The asymptotic approximation of the total transmit power easily follows taking into account that the transmit power of BS $j$ is given by $\frac{1}{N}\sum_{k=1}^{K}\overline{p}_{jk} = \overline{P}_{j}$. This completes the proof.

## APPENDIX D
## PROOF OF COROLLARY 3

We begin by rewriting $(\mathbf{\Gamma}-\mathbf{F})\overline{\mathbf{P}}^{(CoBF)} = \sigma^{2}\mathbf{b}$ as $\mathbf{\Lambda}^{-1}(\mathbf{\Gamma}-\mathbf{F})\overline{\mathbf{P}}^{(CoBF)} = \sigma^{2}\mathbf{a}$ where $\mathbf{\Lambda} \in \mathbb{C}^{L \times L}$ is diagonal with $[\mathbf{\Lambda}]_{l,l} = \eta_{l}$ and $\mathbf{a} \in \mathbb{C}^{L}$ with $a_{l} = \frac{1}{N}\sum_{i=1}^{K}\overline{\lambda}_{li}^{(CoBF)}$. Then, we have that

$$\mathbf{1}^{T}\mathbf{\Lambda}^{-1}(\mathbf{\Gamma}-\mathbf{F})\overline{\mathbf{P}}^{(CoBF)} = \frac{1}{N}\sum_{l=1}^{L}\sum_{i=1}^{K}\overline{\lambda}_{li}^{(CoBF)}\sigma^{2}. \quad (104)$$

Observe now that the $j$th entry of the row vector $\mathbf{1}^{T}\mathbf{\Lambda}^{-1}(\mathbf{\Gamma}-\mathbf{F})$ is equal to unity $\forall j$:

$$\left[\mathbf{1}^{T}\mathbf{\Lambda}^{-1}(\mathbf{\Gamma}-\mathbf{F})\right]_{j} =$$

$$\overset{(a)}{=} \frac{\eta_{j}}{\eta_{j}'} - \frac{1}{N}\frac{1}{\eta_{j}}\sum_{l=1}^{L}\sum_{i=1}^{K}\gamma_{li}\frac{d_{jli}}{d_{lli}}\frac{\eta_{j}}{\eta_{l}}\frac{1}{\left(1+\gamma_{li}\frac{d_{jli}}{d_{lli}}\frac{\eta_{j}}{\eta_{l}}\right)^{2}}$$

$$\overset{(b)}{=} \frac{1}{\eta_{j}}\left[1-\frac{1}{N}\sum_{l=1}^{L}\sum_{i=1}^{K}\frac{\gamma_{li}\frac{d_{jli}}{d_{lli}}\frac{\eta_{j}}{\eta_{l}}}{1+\gamma_{li}\frac{d_{jli}}{d_{lli}}\frac{\eta_{j}}{\eta_{l}}}\right] \overset{(c)}{=} 1$$

where $(a)$ follows from $(24) - (27)$. The expression in $(b)$ is obtained taking into account (87) whereas $(c)$ follows from (18). Plugging the above result into (104) yields the statement of the corollary since the transmit power of BS $j$ is given by $P_{j}^{(CoBF)} = \frac{1}{N}\sum_{k=1}^{K}p_{jk}^{(CoBF)}$.

## APPENDIX E
## PROOF OF THEOREM 3

As done for proving Theorem 1. Assume erroneously for a moment that $\{\lambda_{k}^{*}\}$ are given and independent from $\{\widehat{\mathbf{h}}_{i}\}$. Then, applying [11, Theorem 1] to (33) we conjecture that the ratio $r_{k} = \frac{\overline{\lambda}_{k}^{(CoMP)}}{\lambda_{k}^{*}} = \frac{\gamma_{k}}{\epsilon_{k}}\frac{1}{\lambda_{k}^{*}}$ is equal to unity with $\epsilon_{k}$ being the solution of the following set of equations:

$$\epsilon_{k} = \frac{1}{L}\sum_{l=1}^{L}d_{lk}\mu_{l} = \frac{1}{NL}\text{tr}(\mathbf{\Theta}_{k}\mathbf{T}) \quad (105)$$

with $\mathbf{T} = (\frac{1}{NL}\sum_{i=1}^{KL}\frac{\mathbf{\Theta}_{i}}{\epsilon_{i}}\frac{\gamma_{i}}{1+\gamma_{i}}+\mathbf{I}_{NL})^{-1}$. From (33), using the matrix inversion lemma and (32) one gets

$$\gamma_{k}r_{k} = \frac{1}{NL}\mathbf{z}_{k}^{H}\mathbf{\Theta}_{k}^{1/2^{H}}$$

$$\left(\frac{1}{NL}\sum_{i \neq k}\frac{\overline{\lambda}_{i}}{r_{i}}\mathbf{\Theta}_{i}^{1/2}\mathbf{z}_{i}\mathbf{z}_{i}^{H}\mathbf{\Theta}_{i}^{1/2^{H}}+\mathbf{I}_{NL}\right)^{-1}\mathbf{\Theta}_{k}^{1/2}\mathbf{z}_{k}\overline{\lambda}_{k} \quad (106)$$

where the superscript $^{(CoMP)}$ is omitted for simplicity. Assume that $0 \leq r_{1} \leq r_{2} \leq \cdots \leq r_{KL}$ and that $r_{KL}$ is infinitely often larger than $1 + \ell$. Let us restrict ourselves to such a subsequence. From (106), replacing $r_i$ with $r_{KL}$ and using monotonicity arguments we obtain

$$\gamma_{KL} \leq \frac{1}{NL} \mathbf{z}_{KL}^H \mathbf{\Theta}_{KL}^{1/2^H} \mathbf{B}_k^{-1}(\ell) \mathbf{\Theta}_{KL}^{1/2} \mathbf{z}_{KL} \overline{\lambda}_{KL} \tag{107}$$

with

$$\mathbf{B}_k(\ell) = \frac{1}{NL} \sum_{i \neq k} \overline{\lambda}_i \mathbf{\Theta}_i^{1/2} \mathbf{z}_i \mathbf{z}_i^H \mathbf{\Theta}_i^{1/2^H} + (1 + \ell) \mathbf{I}_{NL}.$$

Applying [11, Theorem 1] yields

$$\frac{1}{NL} \mathbf{z}_{KL}^H \mathbf{\Theta}_{KL}^{1/2^H} \mathbf{B}_k^{-1}(\ell) \mathbf{\Theta}_{KL}^{1/2} \mathbf{z}_{KL} \overline{\lambda}_{KL} \asymp \rho_{KL}(\ell) \tag{108}$$

with $\rho_{KL}(\ell)$ being the unique positive solution to $\rho_k(\ell) = \frac{1}{NL} \mathrm{tr} \left( \mathbf{\Theta}_k \mathbf{T}(\ell) \right)$ with

$$\mathbf{T}(\ell) = \left( \frac{1}{NL} \sum_{i=1}^{KL} \frac{\overline{\lambda}_i \mathbf{\Theta}_i}{1 + \overline{\lambda}_i \rho_i(\ell)} + (1 + \ell) \mathbf{I}_{NL} \right)^{-1}.$$

The proof proceeds as in Appendix A. Omitting the details, we eventually obtain that $1 - \ell \leq r_{KL} \leq 1 + \ell$ from which we may write $\max |r_{KL} - 1| \leq \ell$ for all large values of $KL$ [17]. Taking a countable sequence of $\ell$ going to zero yields $\max |r_{KL} - 1| \to 0$ from which using $r_{KL} = \frac{\overline{\lambda}_{KL}}{\lambda_{KL}^*}$ and assuming $\lim_{KL \to \infty} \sup \gamma_{KL} < \infty$ we obtain $\max |\lambda_{KL}^* - \overline{\lambda}_{KL}| \to 0$ with $\overline{\lambda}_{KL} = \frac{\gamma_{KL}}{\epsilon_{KL}}$. Following the same steps for $k = 1, \ldots, KL - 1$ yields the desired result.

## APPENDIX F
## PROOF OF LEMMA 3

We begin by computing the deterministic equivalent of the interference term of (37). This will be instrumental to obtain the deterministic equivalent of the useful signal power. We start rewriting the denominator of $\mathrm{SINR}_k$ in (37) as $\sum_{i=1, i \neq k}^{KL} p_i \frac{|\hat{\mathbf{h}}_k^H \hat{\mathbf{v}}_i|^2}{||\hat{\mathbf{v}}_i||^2} = \frac{1}{NL} \mathbf{h}_k^H \hat{\mathbf{A}} \left( \frac{1}{NL} \hat{\mathbf{H}}^{[k]} \mathbf{P}^{[k]} \hat{\mathbf{H}}^{[k]H} \right) \hat{\mathbf{A}} \mathbf{h}_k$ with $\hat{\mathbf{A}} = (\frac{1}{NL} \sum_{i=1}^{KL} \overline{\lambda}_i \hat{\mathbf{h}}_i \hat{\mathbf{h}}_i^H + \mathbf{I}_{NL})^{-1}$, $\hat{\mathbf{H}}^{[k]} \triangleq [\hat{\mathbf{h}}_1, \ldots, \hat{\mathbf{h}}_{k-1}, \hat{\mathbf{h}}_{k+1}, \ldots, \hat{\mathbf{h}}_K] \in \mathbb{C}^{N \times (K-1)}$ and $\mathbf{P}^{[k]} \triangleq \mathrm{diag}\{\frac{p_1}{NL ||\hat{\mathbf{v}}_1||^2}, \ldots, \frac{p_{k-1}}{NL ||\hat{\mathbf{v}}_{k-1}||^2}, \frac{p_{k+1}}{NL ||\hat{\mathbf{v}}_{k+1}||^2}, \ldots, \frac{p_K}{NL ||\hat{\mathbf{v}}_K||^2}\}$. Then, mimicking the derivations for CoBF, we have that

$$\frac{1}{NL} \mathbf{h}_k^H \hat{\mathbf{A}} \left( \frac{1}{NL} \hat{\mathbf{H}}^{[k]} \mathbf{P}^{[k]} \hat{\mathbf{H}}^{[k]H} \right) \hat{\mathbf{A}} \mathbf{h}_k \asymp$$
$$\frac{1 - \tau_k^2 \left[ 1 - (1 + \overline{\lambda}_k u)^2 \right]}{(1 + \overline{\lambda}_k u)^2} u' \tag{109}$$

where $u = \frac{1}{NL} \mathrm{tr}(\mathbf{\Theta}_k \hat{\mathbf{A}}^{[k]}) + o(1)$ and $u' = \frac{1}{NL} \mathrm{tr}(\frac{1}{NL} \mathbf{P}^{[k]} \hat{\mathbf{H}}^{[k]} \hat{\mathbf{A}}^{[k]} \mathbf{\Theta}_k \hat{\mathbf{A}}^{[k]} \hat{\mathbf{H}}^{[k]H}) + o(1)$ with $\hat{\mathbf{A}}^{[k]} = (\frac{1}{NL} \sum_{i=1, i \neq k}^{KL} \overline{\lambda}_i \hat{\mathbf{h}}_i \hat{\mathbf{h}}_i^H + \mathbf{I}_{NL})^{-1}$. Observe that $u \asymp \epsilon_k$ (as it follows from Theorem 3) whereas the deterministic equivalent of $u'$ is obtained as follows. Rewrite $\frac{1}{NL} \mathrm{tr}(\frac{1}{NL} \mathbf{P}^{[k]} \hat{\mathbf{H}}^{[k]} \hat{\mathbf{A}}^{[k]} \mathbf{\Theta}_k \hat{\mathbf{A}}^{[k]} \hat{\mathbf{H}}^{[k]H})$ as

$$\frac{1}{NL} \mathrm{tr} \left( \frac{1}{NL} \mathbf{P}^{[k]} \hat{\mathbf{H}}^{[k]} \hat{\mathbf{A}}^{[k]} \mathbf{\Theta}_k \hat{\mathbf{A}}^{[k]} \hat{\mathbf{H}}^{[k]H} \right) =$$
$$\frac{1}{NL} \sum_{i=1, i \neq k}^{KL} p_i \frac{\frac{1}{NL} \hat{\mathbf{h}}_i^H \hat{\mathbf{A}} \mathbf{\Theta}_k \hat{\mathbf{A}} \hat{\mathbf{h}}_i}{\frac{1}{NL} \hat{\mathbf{h}}_i^H \hat{\mathbf{A}}^2 \hat{\mathbf{h}}_i}. \tag{110}$$

Then, observe that

$$\frac{1}{NL} \hat{\mathbf{h}}_i^H \hat{\mathbf{A}} \mathbf{\Theta}_k \hat{\mathbf{A}} \hat{\mathbf{h}}_i \asymp \frac{\frac{1}{NL} \mathrm{tr} \left( \mathbf{\Theta}_i \hat{\mathbf{A}} \mathbf{\Theta}_k \hat{\mathbf{A}} \right)}{\left[ 1 + \frac{1}{NL} \overline{\lambda}_i \mathrm{tr} \left( \mathbf{\Theta}_i \hat{\mathbf{A}} \right) \right]^2}. \tag{111}$$

To compute $\frac{1}{NL} \mathrm{tr}(\mathbf{\Theta}_i \hat{\mathbf{A}} \mathbf{\Theta}_k \hat{\mathbf{A}})$, we may write $\frac{1}{NL} \mathrm{tr}(\mathbf{\Theta}_i \hat{\mathbf{A}} \mathbf{\Theta}_k \hat{\mathbf{A}}) = \frac{1}{NL} \frac{\partial}{\partial z} \mathrm{tr}(\mathbf{\Theta}_i (\hat{\mathbf{A}}^{-1} - z \mathbf{\Theta}_k)^{-1})|_{z=0}$. Observe now that [11, Theorem 1]

$$\frac{1}{NL} \mathrm{tr} \left( \mathbf{\Theta}_i \left( \hat{\mathbf{A}}^{-1} - z \mathbf{\Theta}_k \right)^{-1} \right) \asymp \epsilon_{ik}(z) \tag{112}$$

where $\epsilon_{ik}(z)$ is given by $\epsilon_{ik}(z) = \frac{1}{NL} \mathrm{tr}(\mathbf{\Theta}_i \mathbf{T}_k(z))$ and $\mathbf{T}_k(z)$ is computed as

$$\mathbf{T}_k(z) = \left( \frac{1}{NL} \sum_{n=1}^{KL} \frac{\overline{\lambda}_n \mathbf{\Theta}_n}{1 + \overline{\lambda}_n \epsilon_{nk}(z)} + \mathbf{I}_{NL} - z \mathbf{\Theta}_k \right)^{-1}. \tag{113}$$

By differentiating along $z$, we have $\epsilon'_{ik}(z) = \frac{1}{NL} \mathrm{tr}(\mathbf{\Theta}_i \mathbf{T}'_k(z))$ where $\mathbf{T}'_k(z) = \frac{\partial \mathbf{T}_k(z)}{\partial z}$ is given by $\mathbf{T}'_k(z) = \mathbf{T}_k(z)(\frac{1}{NL} \sum_{n=1}^{KL} \frac{\overline{\lambda}_n^2 \epsilon'_{nk}(z) \mathbf{\Theta}_n}{(1 + \overline{\lambda}_n \epsilon_{nk}(z))^2} + \mathbf{\Theta}_k) \mathbf{T}_k(z)$. Setting $z = 0$ yields

$$\epsilon'_{ik}(0) = \frac{1}{NL} \mathrm{tr} \left( \mathbf{\Theta}_i \mathbf{T}'_k(0) \right) \tag{114}$$

where

$$\mathbf{T}'_k(0) = \mathbf{T} \left( \frac{1}{NL} \sum_{n=1}^{KL} \frac{\overline{\lambda}_n^2 \epsilon'_{nk}(0) \mathbf{\Theta}_n}{(1 + \gamma_n)^2} + \mathbf{\Theta}_k \right) \mathbf{T}. \tag{115}$$

In writing the above result, we have taken into account that $\mathbf{T} = \mathbf{T}_k(0)$, $\epsilon_{nk}(0) = \frac{1}{NL} \mathrm{tr}(\mathbf{\Theta}_n \mathbf{T}_k(0)) = \frac{1}{NL} \mathrm{tr}(\mathbf{\Theta}_n \mathbf{T}) = \epsilon_n$ and $\gamma_n = \overline{\lambda}_n \epsilon_n$. Plugging (115) into (114) and neglecting the functional dependence from $z = 0$, $\epsilon'_k = [\epsilon'_{1k}, \ldots, \epsilon'_{KLk}]^T$ is found as the unique solution of:

$$\epsilon'_{ik} = \frac{1}{NL} \mathrm{tr} \left( \mathbf{\Theta}_i \mathbf{T} \left( \frac{1}{NL} \sum_{n=1}^{KL} \frac{\overline{\lambda}_n^2 \epsilon'_{nk} \mathbf{\Theta}_n}{(1 + \gamma_n)^2} + \mathbf{\Theta}_k \right) \mathbf{T} \right)$$
$$= \frac{1}{NL} \mathrm{tr} \left( \mathbf{\Theta}_i \mathbf{T} \mathbf{\Theta}_k \mathbf{T} \right)$$
$$+ \frac{1}{NL} \sum_{n=1}^{KL} \frac{\overline{\lambda}_n^2 \epsilon'_{nk}}{(1 + \gamma_n)^2} \frac{1}{NL} \mathrm{tr} \left( \mathbf{\Theta}_i \mathbf{T} \mathbf{\Theta}_n \mathbf{T} \right). \tag{116}$$

Observing that $\frac{1}{NL} \mathrm{tr}(\mathbf{\Theta}_i \mathbf{T} \mathbf{\Theta}_k \mathbf{T}) = \frac{1}{L}(\sum_{i=1}^{L} d_{li} d_{lk} \mu_l^2)$ we may rewrite the above system of equations in compact form as $\epsilon'_k = (\mathbf{I}_{KL} - \mathbf{J})^{-1} \mathbf{c}'_k$ with $\mathbf{c}'_k$ and $\mathbf{J}$ being defined in the text (see (38)). On the basis of above results, using $\frac{1}{NL} \overline{\lambda}_i \mathrm{tr}(\mathbf{\Theta}_i \hat{\mathbf{A}}) \asymp \gamma_i$, we eventually obtain that

$$\frac{1}{NL} \hat{\mathbf{h}}_i^H \hat{\mathbf{A}} \mathbf{\Theta}_k \hat{\mathbf{A}} \hat{\mathbf{h}}_i \asymp \frac{\epsilon'_{ik}}{(1 + \gamma_i)^2}. \tag{117}$$

Following similar arguments of above yields

$$\frac{1}{NL} \hat{\mathbf{h}}_i^H \hat{\mathbf{A}}^2 \hat{\mathbf{h}}_i \asymp \frac{\epsilon'_i}{(1 + \gamma_i)^2} \tag{118}$$

with $\boldsymbol{\epsilon}' = [\epsilon_1', \ldots, \epsilon_{KL}']^T = (\mathbf{I}_{KL} - \mathbf{J})^{-1}\mathbf{c}$ where $\mathbf{c} \in \mathbb{C}^{KL}$ has elements $[\mathbf{c}]_i = \frac{1}{L}\sum_{l=1}^{L} d_{li}\mu_i^2$. Putting the results in (117) and (118) together, we have that

$$\frac{1}{NL}\sum_{i=1,i\neq k}^{KL} p_i \frac{\frac{1}{NL}\widehat{\mathbf{h}}_i^H \widehat{\mathbf{A}}\boldsymbol{\Theta}_k\widehat{\mathbf{A}}\widehat{\mathbf{h}}_i}{\frac{1}{NL}\widehat{\mathbf{h}}_i^H \widehat{\mathbf{A}}^2\widehat{\mathbf{h}}_i} \asymp \frac{1}{NL}\sum_{i=1,i\neq k}^{KL} p_i \frac{\epsilon_{ik}'}{\epsilon_i'}. \quad (119)$$

The deterministic equivalent of the numerator of the SINR in (37) is now easily obtained as

$$p_k \frac{|\frac{1}{NL}\mathbf{h}_k^H \widehat{\mathbf{h}}_k|^2}{\frac{1}{NL}\widehat{\mathbf{h}}_k^H \widehat{\mathbf{A}}^2\widehat{\mathbf{h}}_k} \asymp p_k(1-\tau_k^2)\frac{\epsilon_k^2}{\epsilon_k'} \quad (120)$$

since $\frac{1}{NL}\mathbf{h}_k^H \widehat{\mathbf{h}}_k \asymp \sqrt{1-\tau_k^2}\frac{\epsilon_k}{1+\gamma_k}$ and $\frac{1}{NL}\widehat{\mathbf{h}}_k^H \widehat{\mathbf{A}}^2\widehat{\mathbf{h}}_k \asymp \frac{\epsilon_k'}{(1+\gamma_k)^2}$.

## Appendix G
## Proof of Corollary 5

To begin with, we use (46) and (48) to rewrite (51) as follows

$$\overline{p}_{jk} = \overline{\lambda}_{jk}\varsigma_j \frac{\alpha_{jjk}\overline{P}_j + \overline{\sigma}_{jk}^2}{1 - \frac{1}{N}\sum_{i=1}^{K}\frac{\gamma_{ji}^2}{(1+\gamma_{ji})^2}} \quad (121)$$

where $\overline{\sigma}_{jk}^2 = \sum_{l=1,l\neq j}^{L} d_{ljk}\overline{P}_l + \sigma^2$ and the superscript $^{(ScBF)}$ has been omitted for simplicity. Plugging the above expression into $\overline{P}_j = \frac{1}{N}\sum_{k=1}^{K}\overline{p}_{jk}$ and solving with respect to $\overline{P}_j$ one obtain

$$\overline{P}_j = \left(\frac{1 - \frac{1}{N}\sum_{i=1}^{K}\frac{\gamma_{ji}^2}{(1+\gamma_{ji})^2}}{\varsigma_j} - \frac{1}{N}\sum_{k=1}^{K}\overline{\lambda}_{jk}\alpha_{jjk}\right)^{-1}\frac{1}{N}\sum_{k=1}^{K}\overline{\lambda}_{jk}\overline{\sigma}_{jk}^2. \quad (122)$$

Replacing $\alpha_{jjk}$ with its expression in (48) and recalling (46), it follows that (after simple calculus)

$$\frac{1 - \frac{1}{N}\sum_{i=1}^{K}\frac{\gamma_{ji}^2}{(1+\gamma_{ji})^2}}{\varsigma_j} - \frac{1}{N}\sum_{k=1}^{K}\overline{\lambda}_{jk}\alpha_{jjk} = 1. \quad (123)$$

Plugging the above result into (122) completes the proof.

## References

[1] T. Marzetta, "Noncooperative cellular wireless with unlimited numbers of base station antennas," *IEEE Trans. Wireless Commun.*, vol. 9, no. 11, pp. 3590 – 3600, Nov. 2010.

[2] J. Hoydis, S. ten Brink, and M. Debbah, "Massive MIMO in the UL/DL of cellular networks: How many antennas do we need?" *IEEE J. Sel. Areas Commun.*, vol. 31, no. 2, pp. 160 – 171, Feb. 2013.

[3] F. Rusek, D. Persson, B. Lau, E. Larsson, T. Marzetta, O. Edfors, and F. Tufvesson, "Scaling up MIMO: Opportunities and challenges with very large arrays," *IEEE Signal Process. Mag.*, vol. 30, no. 1, pp. 40–60, Jan. 2013.

[4] L. Sanguinetti, A. Moustakas, and M. Debbah, "Interference management in 5G reverse TDD HetNets with wireless backhaul: A large system analysis," *IEEE J. Sel. Areas Commun.*, vol. 33, no. 6, pp. 1187 – 1200, June 2015.

[5] M. Schubert and H. Boche, "Solution of the multiuser downlink beamforming problem with individual SINR constraints," *IEEE Trans. Veh. Tech.*, vol. 53, no. 1, pp. 18–28, Jan. 2004.

[6] A. Wiesel, Y. Eldar, and S. Shamai, "Linear precoding via conic optimization for fixed MIMO receivers," *IEEE Trans. Signal Process.*, vol. 54, no. 1, pp. 161–176, Jan. 2006.

[7] E. Björnson, M. Bengtsson, and B. Ottersten, "Optimal multiuser transmit beamforming: A difficult problem with a simple solution structure [lecture notes]," *IEEE Signal Processing Magazine*, vol. 31, no. 4, pp. 142–148, July 2014.

[8] Y. Chen, S. Zhang, S. Xu, and G. Li, "Fundamental trade-offs on green wireless networks," *IEEE Commun. Mag.*, vol. 49, no. 6, pp. 30–37, June 2011.

[9] D. Gesbert, S. Hanly, H. Huang, S. Shamai Shitz, O. Simeone, and W. Yu, "Multi-cell MIMO cooperative networks: A new look at interference," *IEEE J. Sel. Areas Commun.*, vol. 28, no. 9, Dec. 2010.

[10] H. Dahrouj and W. Yu, "Coordinated beamforming for the multicell multi-antenna wireless system," *IEEE Trans. Wireless Commun.*, vol. 9, no. 5, pp. 1748 – 1759, May 2010.

[11] S. Wagner, R. Couillet, M. Debbah, and D. T. M. Slock, "Large system analysis of linear precoding in correlated MISO broadcast channels under limited feedback," *IEEE Transactions on Information Theory*, vol. 58, no. 7, pp. 4509–4537, July 2012.

[12] L. Sanguinetti, A. Moustakas, E. Björnson, and M. Debbah, "Large system analysis of the energy consumption distribution in multi-user MIMO systems with mobility," *IEEE Trans. Wireless Commun.*, vol. 14, no. 3, pp. 1730 – 1745, March 2015.

[13] S. Lakshminaryana, M. Assaad, and M. Debbah, "Coordinated multi-cell beamforming for massive MIMO: A random matrix approach," *IEEE Trans. Inf. Theory*, vol. 61, no. 6, 2015.

[14] H. Asgharimoghaddam, A. Tolli, and N. Rajatheva, "Decentralizing the optimal multi-cell beamforming via large system analysis," in *Proceedings of the IEEE International Conference on Communications*, Sydney, Australia, June 2014.

[15] J. Zhang, C.-K. Wen, S. Jin, X. Gao, and K.-K. Wong, "Large system analysis of cooperative multi-cell downlink transmission via regularized channel inversion with imperfect CSIT," *IEEE Trans. Wireless Commun.*, vol. 12, no. 10, October 2013.

[16] R. Zakhour and S. Hanly, "Base station cooperation on the downlink: Large system analysis," *IEEE Trans. Inf. Theory*, vol. 58, no. 4, pp. 2079–2106, Apr. 2012.

[17] R. Couillet and M. McKay, "Large dimensional analysis and optimization of robust shrinkage covariance matrix estimators," *Journal of Multivariate Analysis*, vol. 131, no. 0, pp. 99 – 120, 2014.

[18] A. Pennanen, A. Tolli, and M. Latva-aho, "Decentralized coordinated downlink beamforming via primal decomposition," *IEEE Signal Process. Lett.*, vol. 18, no. 11, pp. 647 – 650, Nov 2011.

[19] A. Tolli, H. Pennanen, and P. Komulainen, "Decentralized minimum power multi-cell beamforming with limited backhaul signaling," *IEEE Trans. Wireless Commun.*, vol. 10, no. 2, pp. 570 – 580, Feb. 2011.

[20] L. Sanguinetti, E. Björnson, M. Debbah, and A. Moustakas, "Optimal linear pre-coding in multi-user MIMO systems: A large system analysis," in *Proc. IEEE GLOBECOM*, Austin, USA, 2014.

[21] R. Couillet and M. Debbah, *Random matrix methods for wireless communications*. Cambridge, New York: Cambridge University Press, 2011.

[22] E. Seneta, *Non-negative Matrices and Markov Chains*, 3rd ed. New York, NY: Springer, 2006.

[23] K. Hosseini, W. Yu, and R. Adve, "Large-scale MIMO versus network MIMO for multicell interference mitigation," *IEEE J. Sel. Topics Signal Process.*, vol. 8, no. 5, pp. 930 – 941, Oct 2014.

[24] H. Falconet, L. Sanguinetti, A. Kammoun, and M. Debbah, "Asymptotic analysis of downlink MISO systems over Rician fading channels," in *Proc. IEEE ICASSP*, Shangai, China, 2016.

[25] L. Sanguinetti, A. Kammoun, and M. Debbah, "Asymptotic analysis of multicell massive MIMO over Rician fading channels," in *Proc. IEEE GlobeCom*, Washington, USA, 2016.

[26] R. A. Horn and C. R. Johnson, Eds., *Matrix Analysis*. New York, NY, USA: Cambridge University Press, 1986.